\documentclass[preprint]{aastex}

\newcommand{\yr}{{\rm\,yr}}
\newcommand{\au}{{\rm\,AU}}
\newcommand{\Hoctav}{{\bar{\cal H}_{\rm oct}}}
\newcommand{\dvarpi}{{\varpi_1 - \varpi_2}}

\begin{document}

\title{Secular Evolution of Hierarchical Planetary Systems}
\author{Man Hoi Lee and S. J. Peale}
\affil{Department of Physics, University of California,
       Santa Barbara, CA 93106}
% \email{mhlee@europa.physics.ucsb.edu, peale@io.physics.ucsb.edu}

\begin{abstract}
We investigate the dynamical evolution of coplanar hierarchical
two-planet systems where the ratio of the orbital semimajor axes
$\alpha = a_1/a_2$ is small.
Hierarchical two-planet systems are likely to be ubiquitous among
extrasolar planetary systems.
We show that the orbital parameters obtained from a multiple Kepler
fit to the radial velocity variations of a host star are best
interpreted as Jacobi coordinates and that Jacobi coordinates should
be used in any analyses of hierarchical planetary systems.
An approximate theory that can be applied to coplanar hierarchical
two-planet systems with a wide range of masses and orbital
eccentricities is the octupole-level secular perturbation theory,
which is based on an expansion to order $\alpha^3$ and
orbit-averaging.
It reduces the coplanar problem to one degree of freedom, with $e_1$
(or $e_2$) and $\dvarpi$ as the relevant phase-space variables (where
$e_{1,2}$ are the orbital eccentricities of the inner and outer orbits
and $\varpi_{1,2}$ are the longitudes of periapse).
The octupole equations show that if the ratio of the maximum orbital
angular momenta, $\lambda = L_1/L_2 \approx (m_1/m_2) \alpha^{1/2}$,
for given semimajor axes is approximately equal to a critical value
$\lambda_{\rm crit}$, then libration of $\dvarpi$ about either $0^\circ$ or
$180^\circ$ is almost certain, with possibly large amplitude
variations of both eccentricities.
From a study of the HD~168443 and HD~12661 systems and their variants
using both the octupole theory and direct numerical orbit integrations,
we establish that the octupole theory is highly accurate for systems
with $\alpha \la 0.1$ and reasonably accurate even for systems with
$\alpha$ as large as $1/3$, provided that $\alpha$ is not too close to
a significant mean-motion commensurability or above the stability
boundary.
The HD~168443 system is not in a secular resonance and its $\dvarpi$
circulates.
The HD~12661 system is the first extrasolar planetary system found to
have $\dvarpi$ librating about $180^\circ$.
The secular resonance means that the lines of apsides of the two
orbits are on average anti-aligned, although the amplitude of
libration of $\dvarpi$ is large.
The libration of $\dvarpi$ and the large-amplitude variations of both
eccentricities in the HD~12661 system are consistent with the analytic
results on systems with $\lambda \approx \lambda_{\rm crit}$.
The evolution of the HD~12661 system with the best-fit orbital
parameters and $\sin i = 1$ ($i$ is the inclination of the
orbital plane from the plane of the sky) is affected by the close
proximity to the 11:2 mean-motion commensurability, but small
changes in the orbital period of the outer planet within the
uncertainty can result in configurations that are not affected by
mean-motion commensurabilities.
The stability of the HD~12661 system requires $\sin i > 0.3$.
\end{abstract}
% \keywords{celestial mechanics --- planetary systems ---
%           planets and satellites: general}

\section{INTRODUCTION}

Extrasolar planet searches using high-precision radial velocity
observations to measure the reflex motion of the host stars along the
line of sight have now yielded over $100$ extrasolar
planets.\footnote{
See, e.g.,
\anchor{http://www.obspm.fr/planets}{http://www.obspm.fr/planets}
for a continuously updated catalog.
}
The discoveries include $10$ multiple planet systems with either two
or, in the cases of $\upsilon$~And and 55~Cnc, three detected
planets.
There are also indications that about half of the stars with one known
planet are likely to have additional detectable distant companions
(Fischer et al. 2001).
Many of the known multiple planet systems exhibit interesting
dynamics.
The two planets about GJ~876 are deep in three orbital resonances at
the 2:1 mean-motion commensurability (Laughlin \& Chambers 2001; Lee
\& Peale 2002), and there are possibly two other systems with
mean-motion resonances:
2:1 for the two planets about HD~82943\footnote{
See \anchor{http://obswww.unige.ch/~udry/planet/hd82943syst.html}
{http://obswww.unige.ch/$\sim$udry/planet/hd82943syst.html}.
}
(Go\'zdziewski \& Maciejewski 2001) and 3:1 for the inner two planets
about 55~Cnc (Marcy et al. 2002; Lee \& Peale 2003).
The outer two planets of the $\upsilon$~And system are apparently
locked in a secular resonance with the lines of apsides of the two
orbits being aligned on average (Rivera \& Lissauer 2000; Lissauer \&
Rivera 2001; Chiang, Tabachnik, \& Tremaine 2001), and as we shall
demonstrate in this paper, the two planets of the HD~12661 system are
also locked in a secular resonance, but with the lines of apsides
being anti-aligned on average.
The resonances are often vital in ensuring the long term stability of
the multiple planet systems. 

Although the origin, evolution, and stability of the orbital
configurations of multiple planet systems can usually be analyzed with
direct numerical integrations of the full equations of motion,
a theoretical understanding of the dynamics often requires the
development and application of theories with analytic approximations.
The approximate theories also allow one to explore the parameter space
much more rapidly than direct numerical integrations in the regions
where the approximations are valid.

Because the orbital eccentricities and inclinations of the planets in
the solar system are generally small while the ratios of the orbital
semimajor axes of adjacent planets are generally large ($a_j/a_{j+1}
\ga 0.5 $ except for the Mars-Jupiter pair), the classical
perturbation theory developed for the solar system is based on an
expansion of the disturbing functions that describe the mutual
gravitational interactions of the planets in powers of the
eccentricities and inclinations.
In particular, the Laplace-Lagrange secular solution for the evolution
of a two-planet system that is not affected by mean-motion
commensurabilities is based on retaining just the secular terms (i.e.,
those not involving the mean longitudes) in the disturbing functions
up to second order in the eccentricities and inclinations (see, e.g.,
Murray \& Dermott 1999).
Since high orbital eccentricities are common among extrasolar planets
and the distribution of the orbital eccentricities of the extrasolar
planets in the known multiple planet systems is similar to that of the
single planets (Fischer et al. 2003),
the classical Laplace-Lagrange secular perturbation theory is not in
general adequate for describing the secular evolution of the
extrasolar planetary systems that are not affected by mean-motion
commensurabilities.

On the other hand, many of the known multiple planet systems are
{\it hierarchical} in the sense that the ratio(s) of the semimajor
axes ($a_1/a_2$ for the two-planet systems and $a_1/a_2$ and/or
$a_2/a_3$ for the three-planet systems) are small.
There are $12$ pairs of adjacent planets in the $10$ known multiple
planet systems (two of which have three planets).
Among the $9$ pairs that are not known or suspected to be in
mean-motion resonances,
five have $a_j/a_{j+1} \la 0.1$ and all but one (47~UMa) have
$a_j/a_{j+1} \la 1/3$.
Thus a secular perturbation theory for a two-planet system that is
based on an expansion in $a_1/a_2$ and valid for high eccentricities
should provide an accurate description of the secular evolution of
many extrasolar planetary systems.

It is useful to consider hierarchical two-planet systems in the
context of the general hierarchical triple systems in which a third
body orbits an inner binary on a much wider orbit.
A hierarchical triple system can be treated as two binaries on slowly
perturbed Kepler orbits by using Jacobi coordinates, where the
position of the secondary of mass $m_1$ of the inner binary is
relative to the primary of mass $m_0$ and the position of the third
body of mass $m_2$ is relative to the center of mass of $m_0$ and
$m_1$.
A hierarchical two-planet system is simply a hierarchical triple
system with $m_1$ and $m_2$ much smaller than $m_0$.
In \S~2 we derive the orbital parameters in Jacobi coordinates
obtained by the observers from a two (or more generally multiple)
Kepler fit to the radial velocity variations of a host star and show
that Jacobi coordinates should be used in any analyses of hierarchical
(and possibly other types of) planetary systems.
Star-centered or astrocentric coordinates can introduce significant
high-frequency variations in orbital elements that should be nearly
constant on orbital timescales.
The high-frequency variations can then lead to erroneous sensitivity
of the evolution of the orbital elements to the starting epoch.

Secular perturbation theories based on an expansion in $\alpha =
a_1/a_2$ have been developed for hierarchical triple systems.
The expansion to order $\alpha^2$, called the quadrupole
approximation, was developed by Kozai (1962) and Harrington (1968).
This quadrupole-level secular perturbation theory is successful in
explaining the Kozai mechanism, whereby the perturbations between the
inner binary and the third body can lead to large variations in the
eccentricity of the inner binary and the mutual inclination angle
between the inner and outer binaries if the initial mutual inclination
is sufficiently high.
However, the quadrupole approximation is not adequate for studying
hierarchical two-planet systems.
Although there is no direct information on the mutual inclination
angles in the known hierarchical two-planet systems, the formation of
planets from a common disk of materials surrounding the host star
makes nearly coplanar orbits the most probable configuration.
When the orbits are coplanar, the conservation of total angular
momentum of the system and the secularly constant semimajor axes mean
that the eccentricities of the two orbits are coupled and oscillate
out of phase.
The quadrupole term does not contribute to these eccentricity
oscillations for coplanar orbits.
Marchal (1990), Krymolowski \& Mazeh (1999), and Ford, Kozinsky, \&
Rasio (2000) have extended the approximation to octupole (i.e.,
$\alpha^3$) order.
Blaes, Lee, \& Socrates (2002) have recently applied this
octupole-level secular perturbation theory, with a sign error
corrected (see footnote \ref{signerror} below) and with modifications
to include the effects of general relativistic precession and
gravitational radiation on the inner binary, to study the dynamical
evolution of hierarchical triples of supermassive black holes.
Unlike the quadrupole term, the octupole term does produce
eccentricity oscillations for coplanar orbits.

In this paper we use both the octupole-level secular perturbation
theory and direct numerical orbit integrations to investigate the
dynamical evolution of coplanar hierarchical two-planet systems.
The applicability of the octupole theory is limited by its use of an
expansion in $\alpha$ and an averaging over the inner and outer
orbital motions.
In particular, the orbit averaging eliminates the effects of
mean-motion commensurabilities and the possible development of
instabilities.
We establish the validity and limits of the octupole theory by
comparison with direct numerical orbit integrations.
In \S~3 we summarize the derivation of the octupole theory, compare
the octupole theory to the classical Laplace-Lagrange secular
solution, and deduce some useful results from the octupole equations
analytically.
In \S~4 we study the dynamical evolution of the HD~168443 and HD~12661
systems and their variants.
The HD~168443 system is not in a secular resonance and its secular
resonance variable $\dvarpi$ circulates, where $\varpi_{1,2}$ are the
longitudes of periapse of the inner and outer orbits, respectively.
For the HD~168443 system and a wide variety of systems with $\alpha
\approx 0.1$ (including some for which the octupole theory predicts
rather unusual dynamical behaviors), we show that the octupole results
are in excellent agreement with the direct integration results.
Direct integrations of two-planet systems similar to HD~168443, but
with different initial $a_2$ (and hence different initial $\alpha$),
are used to show that systems with initial $\alpha$ above a critical
value are generally unstable.
As anticipated by the analytic results derived in \S~3, the HD~12661
system is in a secular resonance with $\dvarpi$ librating about
$180^\circ$, and it shows large amplitude variations of both
eccentricities.
We show that the evolution of the HD~12661 system with the best-fit
orbital parameters and $\sin i = 1$ ($i$ is the inclination of
the orbital plane from the plane of the sky) is affected by the
proximity to the 11:2 mean-motion commensurability, but that small
changes in the orbital period of the outer planet within the
uncertainty can result in configurations that are not affected by
mean-motion commensurabilities.
For the latter type of configurations, we show that the octupole
results are in reasonably good agreement with the direct integration
results, even though $\alpha$($\approx 0.32$) is quite large.
We also consider the effects of varying the inclination $i$ and show
that the HD~12661 system is unstable if $\sin i \la 0.3$.
Our conclusions are summarized in \S~5.

\section{JACOBI ORBITAL PARAMETERS FROM MULTIPLE KEPLER FITS TO
         RADIAL VELOCITY OBSERVATIONS}

Except for systems such as GJ~876, where the perturbations between the two
planets are significant on orbital timescales and a dynamical fit to the
stellar radial velocity variations is essential (Laughlin \& Chambers 2001;
Rivera \& Lissauer 2001; Nauenberg 2002),
it is often adequate to fit the radial velocity variations of a star
with two (or more) planets over the time span of the available observations
by assuming that the planets are on unperturbed Kepler orbits.
Many authors have assumed that the orbital parameters obtained from
the multiple Kepler fits are in astrocentric coordinates, but
Lissauer \& Rivera (2001; see also Note Added in Proof of Rivera \&
Lissauer 2000) have pointed out that Jacobi coordinates ``better emulate''
the assumption of unperturbed Kepler orbits.
In this section we derive explicitly the orbital parameters in Jacobi
coordinates obtained from a two-Kepler fit.
It is straightforward to generalize the derivation to an $N$-Kepler fit.
We show that especially for hierarchical systems such as HD~168443, the
use of astrocentric coordinates can introduce erroneous features in the
evolution of the orbital elements.

Let us consider a system consisting of a central star of mass $m_0$, an inner
planet of mass $m_1$, and an outer planet of mass $m_2$, and
use Jacobi coordinates, with $\mbox{\boldmath $r$}_1$ being the position
of $m_1$ relative to $m_0$ and $\mbox{\boldmath $r$}_2$ being the position
of $m_2$ relative to the center of mass of $m_0$ and $m_1$.
We shall refer to the orbit of $m_1$ relative to $m_0$ as the inner
orbit and the orbit of $m_2$ relative to the center of mass of $m_0$ and
$m_1$ as the outer orbit.

If the inner orbit is an unperturbed Kepler orbit,
the line-of-sight (LOS) component of the velocity of $m_1$ relative to
$m_0$ is
\begin{eqnarray}
V_{1,r} &=& - \left[{\dot r}_1 \sin(\omega_1 + f_1) +
                    r_1 {\dot f}_1 \cos(\omega_1 + f_1)\right] \sin i_1
\nonumber \\
        &=& {-2 \pi a_1 \over P_1 \sqrt{1 - e_1^2}}
            \left[\cos(\omega_1 + f_1) + e_1 \cos\omega_1\right] \sin i_1 ,
\end{eqnarray}
where a dot over a symbol denotes $d/dt$ and $a_1$, $e_1$, $i_1$,
$\omega_1$, $f_1$, and
\begin{equation}
P_1 = {2 \pi a_1^{3/2} \over \sqrt{G (m_0 + m_1)}}
\label{period1}
\end{equation}
are, respectively, the semimajor axis, eccentricity, inclination, argument
of periapse, true anomaly, and period of the inner orbit.
In equation (\ref{period1}) $G$ is the gravitational constant.
Note that the reference plane is the plane of the sky, and the planet
approaches the observer at the ascending node, but the radial velocity of
an approaching planet is negative.
Then the LOS component of the velocity of $m_0$ relative to the center of
mass of $m_0$ and $m_1$ is
\begin{equation}
V'_{0,r} = {-m_1 \over m_0 + m_1} V_{1,r}
         = K_1 \left[\cos(\omega_1 + f_1) + e_1 \cos\omega_1\right] ,
\end{equation}
where the amplitude
\begin{equation}
K_1 = \left(2 \pi G \over P_1\right)^{1/3}
      {m_1 \sin i_1 \over (m_0 + m_1)^{2/3}} {1 \over \sqrt{1 - e_1^2}} .
\label{ampl1}
\end{equation}

Similarly, if the outer orbit is an unperturbed Kepler orbit,
the LOS component of the velocity of $m_2$ relative to the center of mass
of $m_0$ and $m_1$ is
\begin{eqnarray}
V_{2,r} = {-2 \pi a_2 \over P_2 \sqrt{1 - e_2^2}}
          \left[\cos(\omega_2 + f_2) + e_2 \cos\omega_2\right] \sin i_2 ,
\end{eqnarray}
where $a_2$, $e_2$, $i_2$, $\omega_2$, $f_2$, and
\begin{equation}
P_2 = {2 \pi a_2^{3/2} \over \sqrt{G (m_0 + m_1 + m_2)}}
\label{period2}
\end{equation}
are, respectively, the semimajor axis, eccentricity, inclination, argument
of periapse, true anomaly, and period of the outer orbit.
Then the LOS component of the velocity of the center of mass of $m_0$ and
$m_1$ relative to the center of mass of the whole system is
\begin{equation}
V'_{01,r} = {-m_2 \over m_0 + m_1 + m_2} V_{2,r}
          = K_2 \left[\cos(\omega_2 + f_2) + e_2 \cos\omega_2\right] ,
\end{equation}
where the amplitude
\begin{equation}
K_2 = \left(2 \pi G \over P_2\right)^{1/3} {m_2 \sin i_2 \over
      (m_0 + m_1 + m_2)^{2/3}} {1 \over \sqrt{1 - e_2^2}} .
\label{ampl2}
\end{equation}

\begin{deluxetable}{lccccc}
\tablecolumns{6}
\tablewidth{0pt}
\tablecaption{Orbital Parameters of the HD~168443 and HD~12661 Planets
\label{table1}}
\tablehead{
\colhead{} & \multicolumn{2}{c}{HD~168443} & \colhead{} & \multicolumn{2}{c}{HD~12661} \\
\cline{2-3} \cline{5-6} \\
\colhead{Parameter} & \colhead{Inner} & \colhead{Outer} & \colhead{} & \colhead{Inner} & \colhead{Outer}
}
\startdata
$P$ (days) & 58.10 & 1770 & & 263.3 & 1444.5\\
$K$ (${\rm m}\,{\rm s}^{-1}$) & 472.7 & 289 & & 74.4 & 27.4\\
$e$ & 0.53 & 0.20 & & 0.35 & 0.20\\
$\omega$ (deg) & 172.9 & 62.9 & & 292.6 & 147.0\\
$T_{\rm peri}$ (JD) & 2450047.58 & 2450250.6 & & 2449943.7 & 2449673.9\\
$m$ ($M_J$) & 7.73 & 17.23 & & 2.30 & 1.57\\
$a$ (AU) & 0.295 & 2.90 & & 0.823 & 2.56\\
\enddata
\tablecomments{The parameters $P$, $K$, $e$, $\omega$, and $T_{\rm peri}$
from two-Kepler fits by Marcy et al. 2001 and D. A. Fischer 2002,
private communication.
The parameters $m$ and $a$ are derived for $\sin i = 1$ and adopted
stellar masses of $1.01 M_\odot$ and $1.07 M_\odot$ for HD~168443 and
HD~12661, respectively.}
\end{deluxetable}

Thus, if the orbits of the planets in Jacobi coordinates are unperturbed
Kepler orbits, the radial velocity of the star $m_0$ is
\begin{eqnarray}
V_r &=& V'_{0,r} + V'_{01,r} + V \nonumber \\
    &=& K_1 \left[\cos(\omega_1 + f_1) + e_1 \cos\omega_1\right] +
        K_2 \left[\cos(\omega_2 + f_2) + e_2 \cos\omega_2\right] + V ,
\label{RV}
\end{eqnarray}
where $V$ is the LOS velocity of the center of mass of the whole
system relative to the observer.
Equation (\ref{RV}) is exactly the formula used by observers in two-Kepler
fits, but with the amplitudes defined in equations (\ref{ampl1}) and
(\ref{ampl2}) and the orbital periods defined in equations (\ref{period1})
and (\ref{period2}).\footnote{
Note that the planetary masses in the equations for the amplitudes and the
orbital periods are the physical masses $m_j$ and not the Jacobi
masses $m_j \sum_{k=0}^{j-1} m_k/\sum_{k=0}^j m_k$ as stated by Lissauer \&
Rivera (2001).
}
Since the true anomaly $f_j$ depends on $P_j$, $e_j$, and the time of
periapse passage $T_{{\rm peri},j}$, a two-Kepler fit directly yields five
parameters $P_j$, $K_j$, $e_j$, $\omega_j$, and $T_{{\rm peri},j}$ for
each orbit.
Table \ref{table1} shows these parameters for the HD~168443 planets from
Marcy et al. (2001) and the HD~12661 planets from D. A. Fischer (2002,
private communication).
If the stellar mass $m_0$ is known, equations (\ref{period1}),
(\ref{ampl1}), (\ref{period2}), and (\ref{ampl2}) can be used to derive
$m_j$ and $a_j$ for assumed $\sin i_j$.
(A common but less accurate practice is to derive $m_j \sin i_j$ and $a_j$
by assuming that $m_1$ and $m_2$ are negligible compared to $m_0$.)
Table \ref{table1} also shows $m_j$ (in units of Jupiter mass $M_J$) and
$a_j$ for $\sin i_j = 1$.
Throughout this paper, we refer to the substellar companions of
HD~168443 as planets, but it should be noted that their $\sin i_j = 1$
(i.e., minimum) masses are about $7.7$ and $17 M_J$, which are near or
above the deuterium-burning limit.

The alternative assumption that the orbits of the planets in
astrocentric coordinates are unperturbed Kepler orbits would also
yield equation (\ref{RV}) for the radial velocity of the host star,
but with $K_j =  [2 \pi G (m_0 + m_j) / P_j]^{1/3} m_j \sin i_j (m_0 +
m_1 + m_2)^{-1} (1 - e_j^2)^{-1/2}$ and $P_j = 2 \pi a_j^{3/2} [G (m_0
+ m_j)]^{-1/2}$.
However, as we show next, while the Jacobi orbits of the planets of a
hierarchical system are nearly Keplerian on orbital timescales,
the astrocentric orbit of the outer planet can deviate significantly
from a Kepler orbit on orbital timescales.

\begin{figure}[t]
\epsscale{1.1}
\plottwo{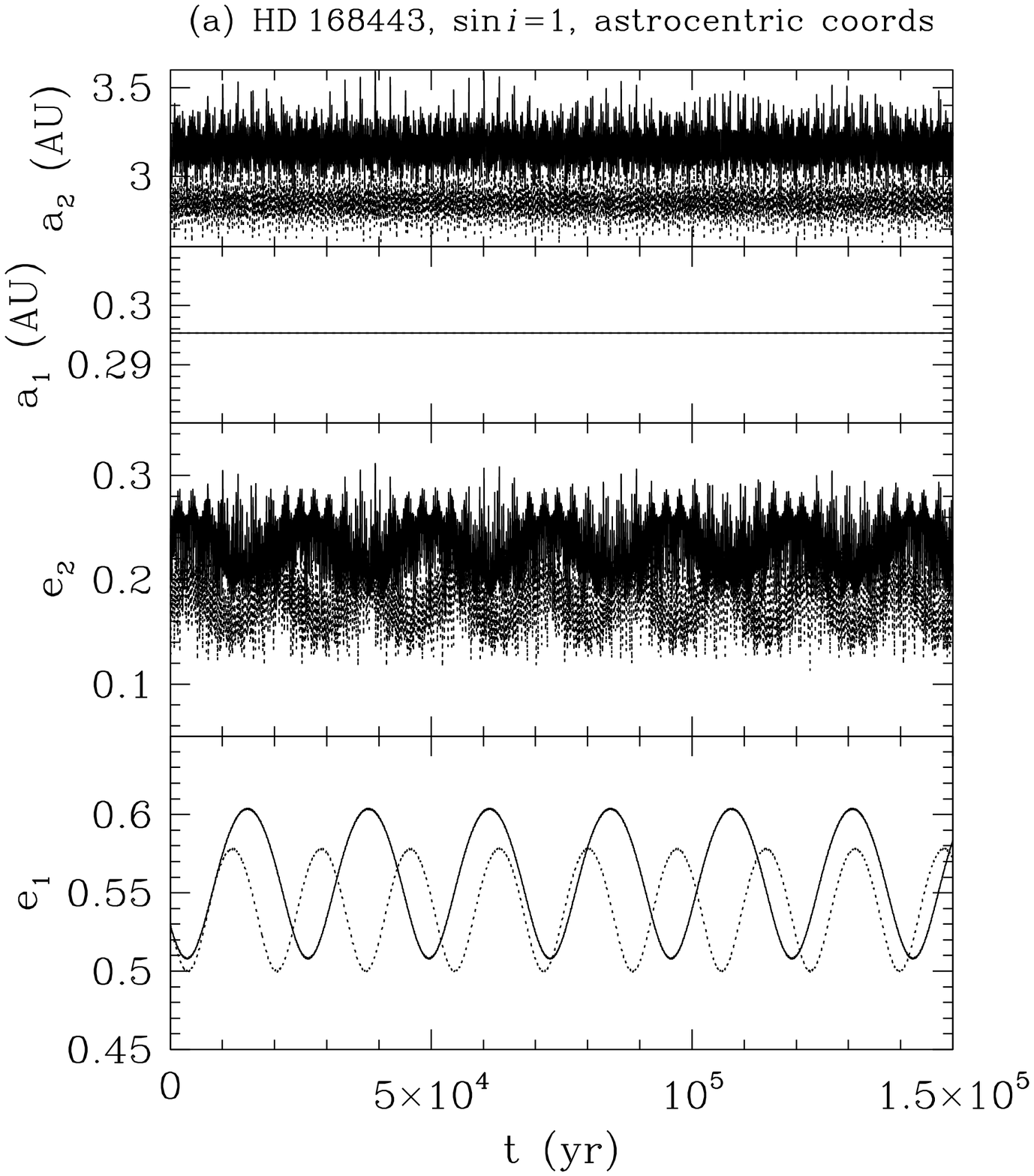}{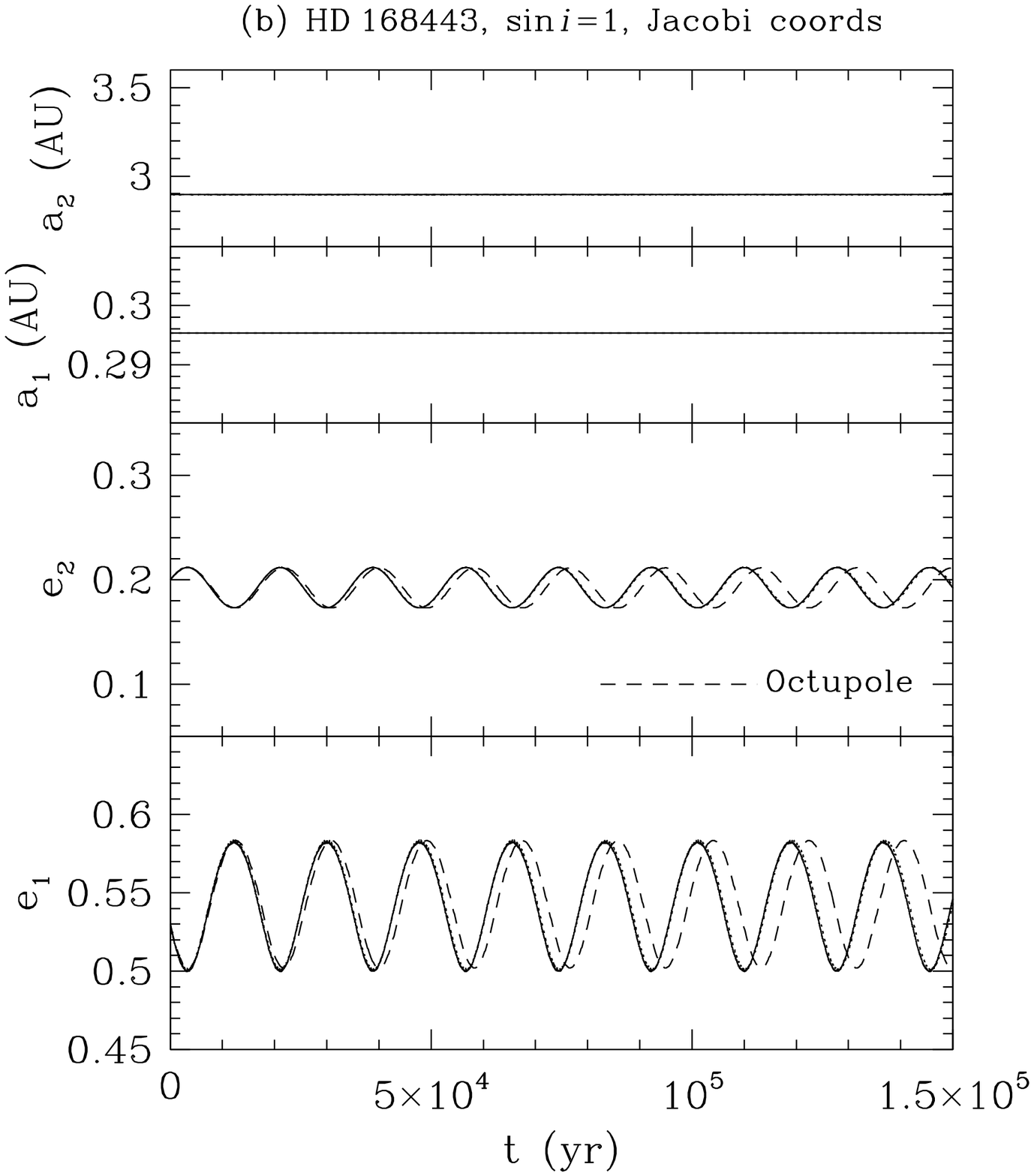}
\caption{
({\it a}) Variations in the orbital semimajor axes, $a_1$ and $a_2$,
and eccentricities, $e_1$ and $e_2$, of the HD~168443 planets with
$\sin i = 1$, if we assume that the best-fit orbital
parameters obtained from the two-Kepler fit are in astrocentric
coordinates and plot the astrocentric $a_j$ and $e_j$.
The solid and dotted lines are from direct numerical orbit
integrations starting at $T_{{\rm peri},1}$ and $T_{{\rm peri},2}$ of
Table 1, respectively.
({\it b}) Same as ({\it a}), but we interpret the
best-fit orbital parameters obtained from the two-Kepler fit as
orbital parameters in Jacobi coordinates and plot the Jacobi $a_j$ and
$e_j$.
The solid and dotted lines are almost indistinguishable.
The dashed lines in the lower two panels of ({\it b}) are from the
octupole-level secular perturbation theory.
\label{figure1}}
\end{figure}

Figure 1{\it a} shows the variations in the semimajor axes and
eccentricities of the HD~168443 planets if we assume that the
orbital parameters obtained from the two-Kepler fit are in astrocentric
coordinates and plot the astrocentric $a_j$ and $e_j$.
The planets are assumed to be on coplanar orbits with $\sin i = 1$
(note that coplanar orbits have the same inclinations, $i_1 = i_2 =
i$, and the same longitudes of ascending node, $\Omega_2 = \Omega_1$,
referenced to the plane of the sky).
The solid and dotted lines are from direct numerical orbit integrations
starting at $T_{{\rm peri},1}$ and $T_{{\rm peri},2}$ of Table \ref{table1},
respectively.
The direct integrations were performed using the Wisdom-Holman (1991)
integrator contained in the SWIFT\footnote{
See \anchor{http://www.boulder.swri.edu/~hal/swift.html}
{http://www.boulder.swri.edu/$\sim$hal/swift.html}.
}
software package, with input and output in astrocentric coordinates.
Figure 1{\it b} is similar to Figure 1{\it a}, but it shows the
variations in the Jacobi $a_j$ and $e_j$, with the initial orbital
parameters also in Jacobi coordinates.
The direct integrations were performed using the modified
Wisdom-Holman integrator described in \S~4.
In astrocentric coordinates (Fig.~1{\it a}), the evolution of the orbital
elements is sensitive to the starting epoch; and
while $a_1$ is nearly constant and $e_1$ oscillates only on the secular
timescale, both $a_2$ and $e_2$ show significant fluctuations on the
orbital timescales (note that we have reduced overcrowding in
Fig.~1{\it a} by using a sampling interval of $50 \yr$, which is long
compared to the orbital timescales).
Plots of the variations in the orbital elements of the HD~168443 planets
in Marcy et al. (2001), Nagasawa, Lin, \& Ida (2003), and Udry, Mayor, \&
Queloz (2003, from calculations by W. Benz) show similar behaviors and
are probably due to the use of astrocentric coordinates.
In contrast, in Jacobi coordinates (Fig.~1{\it b}), the evolution is not
sensitive to the starting epoch (the solid and dotted lines in
Fig.~1{\it b} are almost indistinguishable);
both $a_1$ and $a_2$ are nearly constant, while both $e_1$ and $e_2$
oscillate only on the secular timescale, with the maximum in $e_1$
coinciding with the minimum in $e_2$ and vice versa.
As we shall see, these behaviors can be understood with the
octupole-level secular perturbation theory for coplanar hierarchical
two-planet systems, which is also based on Jacobi coordinates.

From the facts that the Jacobi $a_j$ are nearly constant and that the
Jacobi $e_j$ oscillate only on the secular timescale for hierarchical
systems, it is easy to estimate the fractional fluctuation in the
astrocentric $a_2$, which is primarily due to the velocity of the star
$m_0$ relative to the center of mass of $m_0$ and $m_1$.
It is approximately
$4 (m_1/m_0) [(a_2/a_1) (1+e_{1,{\rm max}})/(1-e_{1,{\rm max}})]^{1/2}$
when the Jacobi $e_1$ is at its maximum $e_{1,{\rm max}}$ (and $e_2$ is at
its minimum, which is assumed to be small and is neglected).
The fluctuations in the astrocentric $a_2$ in Figure 1{\it a} are in
good agreement with this estimate.
Note that the fractional fluctuation is larger for smaller $a_1/a_2$,
i.e., for systems that are more hierarchical.

In their study of the three-planet $\upsilon$~And system, Rivera \&
Lissauer (2000) have also reported that the use of Jacobi coordinates
eliminates the high-frequency variations in the semimajor axes and
eccentricities of the outer two planets and reduces the sensitivity of
the evolution to the initial epoch.
It is clear that Jacobi coordinates should be used in the analysis of
hierarchical systems such as HD~168443 (where $a_1/a_2 \approx 0.10$)
and $\upsilon$~And (where $a_1/a_2 \approx 0.071$ and $a_2/a_3 \approx
0.33$).
It is likely that Jacobi coordinates should also be used in the
analysis of other types of planetary systems, especially those for
which multiple Kepler fits are adequate.

\section{OCTUPOLE-LEVEL SECULAR PERTURBATION THEORY}

As we mentioned in \S~1, an approximate theory that describes the
secular evolution of coplanar hierarchical two-planet systems in
Jacobi coordinates, such as that shown in Figure 1{\it b} for the
HD~168443 system, is the octupole-level secular perturbation theory.
In this section we summarize the derivation of this octupole theory,
compare it to the classical Laplace-Lagrange secular perturbation
theory, and deduce some results from the octupole equations
analytically.

\subsection{Equations}

As in \S 2,
we consider a system consisting of a central star of mass $m_0$, an inner
planet of mass $m_1$, and an outer planet of mass $m_2$, and
use Jacobi coordinates, with $\mbox{\boldmath $r$}_1$ being the position
of $m_1$ relative to $m_0$ and $\mbox{\boldmath $r$}_2$ being the position
of $m_2$ relative to the center of mass of $m_0$ and $m_1$.
The Hamiltonian of this system is
\begin{equation}
{\cal H} = - {G m_0 m_1 \over 2 a_1} - {G (m_0 + m_1) m_2 \over 2 a_2}
           - G m_0 m_2 \left({1 \over r_{02}} - {1 \over r_2}\right)
           - G m_1 m_2 \left({1 \over r_{12}} - {1 \over r_2}\right) ,
\label{ham}
\end{equation}
where $a_j$ is the osculating semimajor axis of the $j$th orbit (with
$j = 1$ and $2$ for the inner and outer orbits, respectively), and $r_{k2}$
is the distance between $m_k$ and $m_2$.
With $r_1 < r_2$, both $1/r_{02}$ and $1/r_{12}$ can be expanded in powers
of $r_1/r_2$, leading to the expanded Hamiltonian
\begin{equation}
{\cal H} = - {G m_0 m_1 \over 2 a_1} - {G (m_0 + m_1) m_2 \over 2 a_2}
           - {G \over a_2} \sum_{k=2}^\infty \alpha^k M_k
             \left(r_1 \over a_1\right)^k \left(a_2 \over r_2\right)^{k+1}
             P_k(\cos\Phi) ,
\label{hamexp}
\end{equation}
where $\alpha = a_1/a_2$,
\begin{equation}
M_k = m_0 m_1 m_2 {m_0^{k-1} - (-m_1)^{k-1} \over (m_0 + m_1)^k} ,
\end{equation}
$P_k$ is the Legendre polynomial of degree $k$, and $\Phi$ is the angle
between $\mbox{\boldmath $r$}_1$ and $\mbox{\boldmath $r$}_2$.
The first two terms in equations (\ref{ham}) and (\ref{hamexp})
represent the independent Kepler motions of the inner and outer
orbits, while the remaining terms represent the perturbations to the
Kepler motions.
For hierarchical systems with $r_1 \ll r_2$, the Jacobi decomposition leads
to two slowly perturbed Kepler orbits, even if $m_1$ and $m_2$ are not much
smaller than $m_0$.

In the general case where mutually inclined orbits are allowed, it is
convenient to use the Delaunay variables with the invariable plane as
the reference plane.
The coordinates are the mean anomalies $l_j$, the arguments of periapse
$g_j = \omega_j$, and the longitudes of ascending node $h_j = \Omega_j$,
and the conjugate momenta are
\begin{eqnarray}
L_1 & = & {m_0 m_1 \over m_0 + m_1} \sqrt{G (m_0 + m_1) a_1} , \\
L_2 & = & {(m_0 + m_1) m_2 \over m_0 + m_1 + m_2}
          \sqrt{G (m_0 + m_1 + m_2) a_2} , \\
G_j & = & L_j \sqrt{1 - e_j^2} , \\
H_j & = & G_j \cos i_j ,
\end{eqnarray}
where $e_j$ and $i_j$ are the eccentricities and inclinations of the
orbits.
The momenta $L_j$, $G_j$, and $H_j$ are, respectively, the magnitude of the
maximum possible angular momentum (if the orbit were circular), the
magnitude of the angular momentum, and the $z$-component of the angular
momentum of the $j$th orbit.
By expressing $\cos\Phi = \mbox{\boldmath $r$}_1 \mbox{\boldmath $\cdot$}
\mbox{\boldmath $r$}_2 / (r_1 r_2)$ in terms of the Delaunay variables,
it is easy to show that the Hamiltonian contains $h_1$ and $h_2$ only in
the combination $h_1 - h_2$, and $h_1 - h_2 = 180^\circ$ if the invariable
plane is used as the reference plane.
Constant $h_1 - h_2$ and constant total angular momentum $H_1 + H_2$ mean
$h_j$ and $H_j$ can be eliminated from the problem --- usually referred to
as the elimination of nodes --- and the system is reduced from $6$ to $4$
degrees of freedom (see, e.g., Marchal 1990).

Kozai (1962) studied the secular evolution of hierarchical triple systems
with $m_1 \ll m_2 \ll m_0$ and $e_2 = 0$ by retaining just the lowest-order
($\alpha^2$) quadrupole term in the series in equation (\ref{hamexp})
and averaging the Hamiltonian over the inner and outer orbital motions.
Harrington (1968) generalized the quadrupole analysis to general
hierarchical triple systems.
Marchal (1990), Krymolowski \& Mazeh (1999), and Ford et al. (2000)
extended the approximation by retaining also the octupole term of order
$\alpha^3$ in the series in equation (\ref{hamexp}) and derived an
octupole-level orbit-averaged Hamiltonian.
The approach used in these studies to average the Hamiltonian is the von
Zeipel method, which involves the determination of a canonical
transformation such that the transformed Hamiltonian is independent of the
transformed mean anomalies $l_1$ and $l_2$, and such that the original and
transformed variables differ only by high frequency terms that are
multiplied by powers of the small parameter $\alpha$.
Since the transformed Hamiltonian is independent of the transformed $l_j$,
the transformed $L_j$ and the corresponding semimajor axes are constant.

In this paper we focus on hierarchical two-planet systems with
coplanar orbits, as the formation of planets from a common disk of
materials surrounding the host star makes nearly coplanar orbits the
most probable configuration.
In the limit of coplanar orbits ($i_1 = i_2 = 0^\circ$), the
Delaunay variables described above are not well-defined, since there are no
ascending nodes with respect to the invariable plane, but the longitudes of
periapse $\varpi_j$ are well-defined.
Thus, for coplanar orbits, it is convenient to use the canonical
variables $l_j$, $\varpi_j$, $L_j$, and $G_j$.
In the limit of coplanar orbits, by noting that the expression for the
angle $\varphi$ between the directions of periapse reduces to
$\cos\varphi = -\cos(g_1 - g_2) = \cos(g_1 - g_2 + 180^\circ) =
\cos(\dvarpi)$ (since $h_1 - h_2 = 180^\circ$), the doubly
averaged octupole-level Hamiltonian derived by Marchal (1990), Krymolowski
\& Mazeh (1999), and Ford et al. (2000) can be written as\footnote{
There is a sign error common to both Krymolowski \& Mazeh (1999) and Ford
et al. (2000).
We follow Ford et al. and denote the coefficients of the quadrupole and
octupole terms by $C_2$ and $C_3$, respectively.
Krymolowski \& Mazeh denote these coefficients by $C_1$ and $C_2$,
respectively.
In the averaged Hamiltonian and the subsequent equations of motion of Ford
et al., the coefficient $C_3$ should be replaced by $-C_3$.
Similarly, in the averaged Hamiltonian and the subsequent equations of
motion of Krymolowski \& Mazeh, the coefficient $C_2$ should be replaced by
$-C_2$.
\label{signerror}
}
\begin{equation}
\Hoctav = - {G^2 m_0^3 m_1^3 \over 2 (m_0 + m_1) L_1^2}
          - {G^2 (m_0 + m_1)^3 m_2^3 \over 2 (m_0 + m_1 + m_2) L_2^2}
          - 2 C_2 (2 + 3 e_1^2)
          + C_3 e_1 e_2 (4 + 3 e_1^2) \cos(\dvarpi) ,
\label{Hoctav}
\end{equation}
where
\begin{eqnarray}
C_2 & = & {1 \over 16} {G^2 (m_0 + m_1)^7 m_2^7 \over
                        (m_0 + m_1 + m_2)^3 (m_0 m_1)^3}
                       {L_1^4 \over L_2^3 G_2^3} , \\
& & \nonumber \\
C_3 & = & {15 \over 64} {G^2 (m_0 + m_1)^9 m_2^9 (m_0 - m_1) \over
                         (m_0 + m_1 + m_2)^4 (m_0 m_1)^5}
                        {L_1^6 \over L_2^3 G_2^5} ,
\end{eqnarray}
and
\begin{equation}
e_j = \sqrt{1 - \left(G_j \over L_j\right)^2} .
\end{equation}
As in Ford et al. (2000), we do not include in equation (\ref{Hoctav})
terms of order $\alpha^{7/2}$ induced by the canonical transformation of
the von Zeipel method,
which were included partially by Marchal (1990) and fully by Krymolowski \&
Mazeh (1999).

As we mentioned above, the orbit averaging eliminates $l_j$ from $\Hoctav$,
and hence $L_j$ (and $a_j$) are constant.
Furthermore, $\Hoctav$ contains $\varpi_1$ and $\varpi_2$ only in the
combination $\dvarpi$, and
choosing this difference as a new variable in a canonical transformation
simply reveals that the total angular momentum $G_1 + G_2$ is an integral
of motion.
Thus, in the octupole approximation, the coplanar problem is reduced
to one degree of freedom with the last two terms in equation (\ref{Hoctav})
as an integral of motion.
We can also see from equation (\ref{Hoctav}) that the quadrupole
approximation is not adequate for studying coplanar systems.
If we drop the octupole term with coefficient $C_3$ in equation
(\ref{Hoctav}), the Hamiltonian is independent of $\varpi_1$ and $\varpi_2$,
meaning that both $e_1$ and $e_2$ are constant, which is not generally true.
(For mutually inclined orbits, only $e_2$ is constant in the quadrupole
approximation; Harrington 1968.)

From the equations of motion
\begin{equation}
{d \varpi_j \over d t} = {\partial \Hoctav \over \partial G_j}
\qquad {\rm and} \qquad
{d G_j \over d t} = - {\partial \Hoctav \over \partial \varpi_j} ,
\end{equation}
we obtain the following equations for the variation of $e_j$ and $\varpi_j$:
\begin{eqnarray}
{d e_1 \over d t} & = & -A_{12} e_2 {(1 - e_1^2)^{1/2}
                                     \left(1 + {3 \over 4} e_1^2\right)
                                     \over (1 - e_2^2)^{5/2}}
                         \sin(\dvarpi) , \label{de1dt} \\
& & \nonumber \\
{d e_2 \over d t} & = &  A_{21} e_1 {\left(1 + {3 \over 4} e_1^2\right)
                                     \over (1 - e_2^2)^2}
                         \sin(\dvarpi) , \label{de2dt} \\
& & \nonumber \\
{d\varpi_1 \over d t}
& = & A_{11} {(1 - e_1^2)^{1/2} \over (1 - e_2^2)^{3/2}}
      - A_{12} \left(e_2 \over e_1\right)
        {(1 - e_1^2)^{1/2} \left(1 + {9 \over 4} e_1^2\right)
         \over (1 - e_2^2)^{5/2}} \cos(\dvarpi) ,
\label{dw1dt} \\
& & \nonumber \\
{d\varpi_2 \over d t}
& = & A_{22} {\left(1 + {3 \over 2} e_1^2\right) \over (1 - e_2^2)^2}
      - A_{21} \left(e_1 \over e_2\right)
        {(1 + 4 e_2^2) \left(1 + {3 \over 4} e_1^2\right)
         \over (1 - e_2^2)^3} \cos(\dvarpi) ,
\label{dw2dt}
\end{eqnarray}
where the constant coefficients
\begin{mathletters}
\begin{eqnarray}
A_{12} &=& {4 C_3 \over L_1} \left(G_2 \over L_2\right)^5
        =  {15 \over 16} n_1 \left(m_2 \over m_0 + m_1\right)
           \left(m_0 - m_1 \over m_0 + m_1\right) \alpha^4 , \\
& & \nonumber \\
A_{21} &=& {4 C_3 \over L_2} \left(G_2 \over L_2\right)^5
        =  {15 \over 16} n_2 {m_0 m_1 \over (m_0 + m_1)^2}
           \left(m_0 - m_1 \over m_0 + m_1\right) \alpha^3 , \\
& & \nonumber \\
A_{11} &=& {12 C_2 \over L_1} \left(G_2 \over L_2\right)^3
        =  {3 \over 4} n_1 \left(m_2 \over m_0 + m_1\right) \alpha^3 , \\
& & \nonumber \\
A_{22} &=& {12 C_2 \over L_2} \left(G_2 \over L_2\right)^3
        =  {3 \over 4} n_2 {m_0 m_1 \over (m_0 + m_1)^2} \alpha^2 ,
\end{eqnarray}
\end{mathletters}with $\alpha = a_1/a_2$ and the mean motions
\begin{equation}
n_1 = \left[G(m_0 + m_1)/a_1^3\right]^{1/2} \qquad {\rm and} \qquad
n_2 = \left[G(m_0 + m_1 + m_2)/a_2^3\right]^{1/2}.
\end{equation}
Consistent with the fact that the system can be reduced to one degree of
freedom, either equation (\ref{de1dt}) or (\ref{de2dt}) is redundant,
since $e_1$ and $e_2$ are related by the conservation of total angular
momentum, and equations (\ref{dw1dt}) and (\ref{dw2dt}) can be combined as
\begin{eqnarray}
{d(\dvarpi) \over d t}
& = & A_{11} \left[{(1 - e_1^2)^{1/2} \over (1 - e_2^2)^{3/2}} -
                   \left(L_1 \over L_2\right)
                   {\left(1 + {3 \over 2} e_1^2\right)
                    \over (1 - e_2^2)^2}\right] \nonumber \\
&   & \nonumber \\
&   & - A_{12} \left[\left(e_2 \over e_1\right)
                     {(1 - e_1^2)^{1/2} \left(1 + {9 \over 4} e_1^2\right)
                      \over (1 - e_2^2)^{5/2}}
                     - \left(L_1 \over L_2\right) \left(e_1 \over e_2\right)
                     {(1 + 4 e_2^2) \left(1 + {3 \over 4} e_1^2\right)
                      \over (1 - e_2^2)^3}\right] \nonumber \\
&   & \nonumber \\
&   & \qquad \times \cos(\dvarpi) .
\label{dw12dt}
\end{eqnarray}

If we relax the assumption of coplanar orbits and allow a small mutual
inclination angle $i_{\rm mu}$ between the inner and outer orbits,
the number of degrees of freedom is increased from one to two, but an
expansion of the orbit-averaged octupole-level Hamiltonian of Marchal
(1990), Krymolowski \& Mazeh (1999), and Ford et al. (2000) in powers
of $i_{\rm mu}$ shows that equations (\ref{de1dt})--(\ref{dw2dt}) are
only modified by terms of order $i_{\rm mu}^2$ and higher.
One of the additional terms for $de_1/dt$ (which has an octupole term
only in the coplanar limit; see eq.~[\ref{de1dt}]) is a quadrupole
term with coefficient $C_2 i_{\rm mu}^2$.
Thus we expect a small mutual inclination angle $i_{\rm mu} \ll
\alpha^{1/2}$ to have little effect on the secular evolution of $e_j$
and $\varpi_j$ of a hierarchical system.
This is confirmed by direct numerical orbit integrations in \S~4.

It is important to note that the octupole-level secular perturbation
theory is derived under the assumptions that there are no mean-motion
commensurabilities and that $\alpha$ (or more precisely $r_1/r_2$)
$\ll 1$.
As we shall see in \S~4, close proximity to even a rather high order
commensurability (e.g., 11:2) can affect the evolution of a system.
We shall also see in \S~4 that although there is a limit to how large
$\alpha$ can be for a system to be stable, the octupole theory can
provide a reasonable description of the secular evolution for $\alpha$ as
large as $1/3$ if the planets are not too massive.

\subsection{Comparison with Classical Laplace-Lagrange Secular
            Perturbation Theory}

As we mentioned in \S~1, the classical secular perturbation theory
developed for the solar system is based on an expansion in the
eccentricities and inclinations.
While it is valid to all orders in the ratio of the semimajor axes
$\alpha$, it can be applied only to systems with small planetary
masses on nearly circular and nearly coplanar orbits.
In contrast, the octupole-level secular perturbation theory is based
on an expansion in $\alpha$, and it can be applied to hierarchical
systems with a wide range of masses, eccentricities and, in its most
general form, inclinations.
We now show that the octupole equations (\ref{de1dt})--(\ref{dw2dt}) agree
with the equations for the classical Laplace-Lagrange secular solution
for coplanar systems where the planetary masses, eccentricities, and
$\alpha$ are all small.

To the lowest order in the eccentricities,
equations (\ref{de1dt})--(\ref{dw2dt}) reduce to
\begin{eqnarray}
{d e_1 \over d t} & = & -A_{12} e_2 \sin(\dvarpi) ,
\label{de1dtII} \\
{d e_2 \over d t} & = &  A_{21} e_1 \sin(\dvarpi) , \\
{d\varpi_1 \over d t}
& = & A_{11} - A_{12} \left(e_2 \over e_1\right) \cos(\dvarpi) , \\
{d\varpi_2 \over d t}
& = & A_{22} - A_{21} \left(e_1 \over e_2\right) \cos(\dvarpi) ,
\label{dw2dtII}
\end{eqnarray}
where $A_{12} = (15/16) (m_2/m_0) n_1 \alpha^4$, $A_{21} = (15/16)
(m_1/m_0) n_2 \alpha^3$, $A_{11} = (3/4) (m_2/m_0) n_1 \alpha^3$, and
$A_{22} = (3/4) (m_1/m_0) n_2 \alpha^2$ in the limit $m_1, m_2 \ll m_0$.

The classical Laplace-Lagrange secular solution is based on retaining
just the secular terms in the disturbing functions up to second order
in the eccentricities and inclinations.
For a coplanar planetary system, the equations for the variation of $e_j$
and $\varpi_j$ (see, e.g., Murray \& Dermott 1999) are of the same form as
equations (\ref{de1dtII})--(\ref{dw2dtII}),
but with the $A_{jk}$ replaced by
\begin{mathletters}
\begin{eqnarray}
A_{12}' &=& {1 \over 4} n_1 {m_2 \over m_0 + m_1} \alpha^2
            b^{(2)}_{3/2}(\alpha) , \\
A_{21}' &=& {1 \over 4} n_2 {m_1 \over m_0 + m_2} \alpha
            b^{(2)}_{3/2}(\alpha) , \\
A_{11}' &=& {1 \over 4} n_1 {m_2 \over m_0 + m_1} \alpha^2
            b^{(1)}_{3/2}(\alpha) , \\
A_{22}' &=& {1 \over 4} n_2 {m_1 \over m_0 + m_2} \alpha
            b^{(1)}_{3/2}(\alpha) ,
\end{eqnarray}
\end{mathletters}where $b^{(j)}_{3/2}(\alpha)$ is the Laplace coefficient.
(It should be noted that the classical theory uses astrocentric coordinates
instead of Jacobi coordinates, but that the distinction between the two
sets of coordinates vanishes in the limit $m_1/m_0 \ll \alpha^{1/2}/4$;
see \S 2.)
Since $b^{(2)}_{3/2}(\alpha) = (15/4) \alpha^2$ and $b^{(1)}_{3/2}(\alpha) =
3 \alpha$ to the lowest order in $\alpha$, the classical and octupole
secular perturbation equations are identical in the limit of small
planetary masses, eccentricities, and $\alpha$.

\subsection{Some Analytic Results}

Although it does not appear that the octupole-level secular perturbation
equations can be solved analytically, some useful results can be deduced
analytically.
We begin by rewriting equations (\ref{de1dt}), (\ref{de2dt}), and
(\ref{dw12dt}) as
\begin{eqnarray}
{d e_1 \over d \tau} & = & -\beta e_2 {(1 - e_1^2)^{1/2}
                                       \left(1 + {3 \over 4} e_1^2\right)
                                       \over (1 - e_2^2)^{5/2}}
                           \sin(\dvarpi) , \label{de1dtau} \\
& & \nonumber \\
{d e_2 \over d \tau} & = & \beta \lambda e_1
                           {\left(1 + {3 \over 4} e_1^2\right)
                             \over (1 - e_2^2)^2}
                           \sin(\dvarpi) , \label{de2dtau} \\
& & \nonumber \\
{d(\dvarpi) \over d \tau}
& = & \left[{(1 - e_1^2)^{1/2} \over (1 - e_2^2)^{3/2}} -
            \lambda {\left(1 + {3 \over 2} e_1^2\right)
                     \over (1 - e_2^2)^2}\right] \nonumber \\
&   & \nonumber \\
&   & - \beta \left[\left(e_2 \over e_1\right)
                    {(1 - e_1^2)^{1/2} \left(1 + {9 \over 4} e_1^2\right)
                     \over (1 - e_2^2)^{5/2}}
                    - \lambda \left(e_1 \over e_2\right)
                    {(1 + 4 e_2^2) \left(1 + {3 \over 4} e_1^2\right)
                     \over (1 - e_2^2)^3}\right] \nonumber \\
&   & \nonumber \\
&   & \qquad \times \cos(\dvarpi) , \label{dw12dtau}
\end{eqnarray}
where
\begin{eqnarray}
\beta &=& {A_{12} \over A_{11}} = {A_{21} \over A_{22}}
       =  {5 \over 4} \left(m_0 - m_1 \over m_0 + m_1\right) \alpha , \\
\lambda &=& {A_{22} \over A_{11}} = {A_{21} \over A_{12}}
         =  {L_1 \over L_2} , \\
\tau &=& t/t_e ,
\end{eqnarray}
and
\begin{equation}
t_e = {1 \over A_{11}}
    = {4 \over 3 \alpha^3} \left(m_0 + m_1 \over m_2\right) {1 \over n_1} .
\end{equation}
Recall that either $e_1$ or $e_2$ can be eliminated from the above
equations by the conservation of total angular momentum, which can be
written in dimensionless form as $\gamma = (G_1 + G_2)/(L_1 + L_2) =
[\lambda (1-e_1^2)^{1/2} + (1-e_2^2)^{1/2}]/(\lambda + 1) =$ constant.

We can see from equations (\ref{de1dtau})--(\ref{dw12dtau}) that coplanar
hierarchical systems with the same $\beta$ and $\lambda$ (which are
constant) and the same initial $e_1$, $e_2$, and $\dvarpi$ have
the same trajectory in the phase-space diagram of $e_1$ (or $e_2$) versus
$\dvarpi$ and can differ only in the period of eccentricity
oscillations.
In particular, in the limit $m_1, m_2 \ll m_0$, since $\beta \approx
(5/4) \alpha$, $\lambda \approx (m_1/m_2) \alpha^{1/2}$, and $t_e \approx
4 (m_0/m_2)/(3 \alpha^3 n_1)$,
systems with the same $\alpha$, $m_1/m_2$, and initial $e_1$, $e_2$ and
$\dvarpi$, but different $m_0$, $m_2$ and/or $a_1$, differ
only in the period of eccentricity oscillations, which is proportional to
$(m_0/m_2)/n_1$.
As we discussed in \S~2, a two-Kepler fit yields $e_j$, $\omega_j$, and
in the limit $m_1, m_2 \ll m_0$, $m_j \sin i_j$ and $a_j$, but not the
inclinations $i_j$ of the orbital planes to the plane of the sky.
(Hereafter, all inclinations are relative to the plane of the sky and
not to the invariable plane.)
If we assume that the orbits are coplanar, the above results mean
that the trajectory in $e_1$ (or $e_2$) versus $\dvarpi$ and
the amplitudes of eccentricity oscillations should be independent of the
unknown inclination $i$ of both orbits, and that the period of eccentricity
oscillations should be proportional to $\sin i$.

The phase-space structure of the secular evolution of coplanar hierarchical
two-planet systems can be understood by plotting the trajectories of systems
with the same $\beta$, $\lambda$, and $\gamma$ in a diagram of $e_1$
(or $e_2$) versus $\dvarpi$ (see, e.g., Figs.~2 and 6 below).
The fixed points in the phase-space diagram can be found by solving
$de_1/d\tau = de_2/d\tau = d(\dvarpi)/d\tau = 0$.
It is clear from equations (\ref{de1dtau}) and (\ref{de2dtau}) that
$de_1/d\tau = de_2/d\tau = 0$ requires $\sin(\dvarpi) = 0$ or
$\dvarpi = 0^\circ$ or $180^\circ$.
As we shall see, there are usually two libration islands, one about a
fixed point at $\dvarpi = 0^\circ$ and another about a fixed
point at $\dvarpi = 180^\circ$;
additional fixed points are possible if the total angular momentum is low.

At $\dvarpi \approx 90^\circ$ and $270^\circ$, since
$\cos(\dvarpi) \approx 0$, equation (\ref{dw12dtau}) reduces to
$d(\varpi_1 - \varpi_2)/d\tau \approx (1 - e_1^2)^{1/2}/(1 - e_2^2)^{3/2} -
\lambda (1 + 3e_1^2/2)/(1 - e_2^2)^2$.
Therefore, to the lowest order in the eccentricities, if $\lambda
\approx 1$, $d(\dvarpi)/d\tau \approx 0$ and $\dvarpi$
should be nearly constant while both $e_1$ and $e_2$ change.
According to equations (\ref{de1dtau}) and (\ref{de2dtau}), $e_1$
should be decreasing (from near its maximum possible value for the
given total angular momentum to near zero) and $e_2$ should be
increasing (from near zero to near its maximum possible value for the
given total angular momentum) at $\dvarpi \approx 90^\circ$, where
$\sin(\dvarpi) \approx 1$, and vice versa at $\dvarpi \approx
270^\circ$, where $\sin(\dvarpi) \approx -1$.
These behaviors are most consistent with a phase space that is
dominated by two large libration islands, one about a fixed point at
$\dvarpi = 0^\circ$ and another about a fixed point at $\dvarpi =
180^\circ$.
Note also that for a given total angular momentum the maximum possible
values of $e_1$ and $e_2$ are comparable if $\lambda \approx 1$.
Finite eccentricities change the condition for $d(\dvarpi)/d\tau
\approx 0$ at $\dvarpi \approx 90^\circ$ and $270^\circ$ to
$\lambda \approx (1 - e_1^2)^{1/2} (1 - e_2^2)^{1/2}/(1 + 3e_1^2/2)$,
which is slightly less than $1$ for moderate eccentricities.
To estimate how much smaller than $1$ the critical value of $\lambda$
is for a given dimensionless total angular momentum $\gamma$, we
substitute into the above condition the value of the eccentricities
when they are equal [$e_1 = e_2 = (1 - \gamma^2)^{1/2}$] and obtain
$\lambda_{\rm crit} = 2 \gamma^2/(5 - 3\gamma^2)$.
Note that $\lambda_{\rm crit} = 1$ is recovered for $\gamma = 1$.
Therefore, large libration islands and large amplitude variations of
{\it both} $e_j$ are likely if $\lambda \approx \lambda_{\rm crit} = 2
\gamma^2/(5 - 3\gamma^2)$.

\section{NUMERICAL RESULTS}

In this section we study the dynamical evolution of the HD~168443 and
HD~12661 systems and their variants and demonstrate the validity and
limits of the octupole-level secular perturbation theory by comparison
with direct numerical orbit integrations.
Except for the direct integrations discussed in \S~4.3, the planets
are assumed to be on coplanar orbits.
Since the octupole-level secular perturbation equations do not involve high
frequency terms, numerical integrations of these equations are rapid.
We integrated equations (\ref{de2dtau}) and (\ref{dw12dtau}), with
$e_1$ found from conservation of total angular momentum, using a
Bulirsch-Stoer integrator.
The direct numerical orbit integrations were performed using a modified
version of the Wisdom-Holman (1991) integrator contained in the SWIFT
software package.
In addition to changing the input and output to Jacobi orbital elements, we
divide the Hamiltonian into a part that describes the Kepler motions
of the inner and outer orbits and a part that describes the
perturbations to the Kepler motions using equation (\ref{ham}) instead
of the division used by Wisdom \& Holman (1991), which moves the term
$G m_1 m_2/r_2$ from the perturbation Hamiltonian to the Kepler
Hamiltonian of the outer orbit.
This modified integrator can handle hierarchical systems where $m_1$
and $m_2$ are not small.

\subsection{HD~168443}

The best-fit orbital parameters of the HD~168443 planets from Marcy et
al. (2001) and the inferred planetary masses and semimajor axes for
$\sin i = 1$ are listed in Table \ref{table1}.
This system has $\alpha = a_1/a_2 = 0.102$, $\beta = 5 (m_0 - m_1)
\alpha/ [4 (m_0 + m_1)] = 0.126$, $\lambda = L_1/L_2 = 0.143$, and
$\gamma = (G_1 + G_2)/(L_1 + L_2) = 0.963$.
Figure 1{\it b} shows the variations in the Jacobi $a_j$ and $e_j$, with
the solid and dotted lines (almost indistinguishable) from direct numerical
orbit integrations starting at $T_{{\rm peri},1}$ and $T_{{\rm peri},2}$ of
Table \ref{table1}, respectively, and the dashed lines in the lower
two panels from the octupole-level secular perturbation theory.
The trajectories in Figure 2 marked by the squares (which denote the
current parameters of the HD~168443 system) are the trajectories of
the $\sin i = 1$ HD~168443 system in the phase-space diagrams of $e_1$
versus $\dvarpi$ (upper panels) and $e_2$ versus $\dvarpi$ (lower
panels), with the left and right panels showing the results from
direct integrations and the octupole theory, respectively (see below
for a discussion of the other trajectories in Fig.~2).
The direct integration results agree with the octupole theory in that the
evolution is not sensitive to the starting epoch and that $a_1$ and $a_2$
are nearly constant.
In addition, the trajectories in the diagram of $e_1$ (or $e_2$) versus
$\dvarpi$ from the direct integrations are in excellent agreement with that
from the octupole theory.
The only noticeable difference is that the period of eccentricity
oscillations predicted by the octupole theory is about $3\%$ longer that
that found in the direct integrations ($\approx 1.8 \times 10^4\yr$).
Krymolowski \& Mazeh (1999) have reported that the oscillation period
predicted by the octupole theory can be improved by the inclusion of the
terms of order $\alpha^{7/2}$ induced by the canonical transformation of
the von Zeipel method and neglected by us.

\begin{figure}[t]
\epsscale{0.7}
\plotone{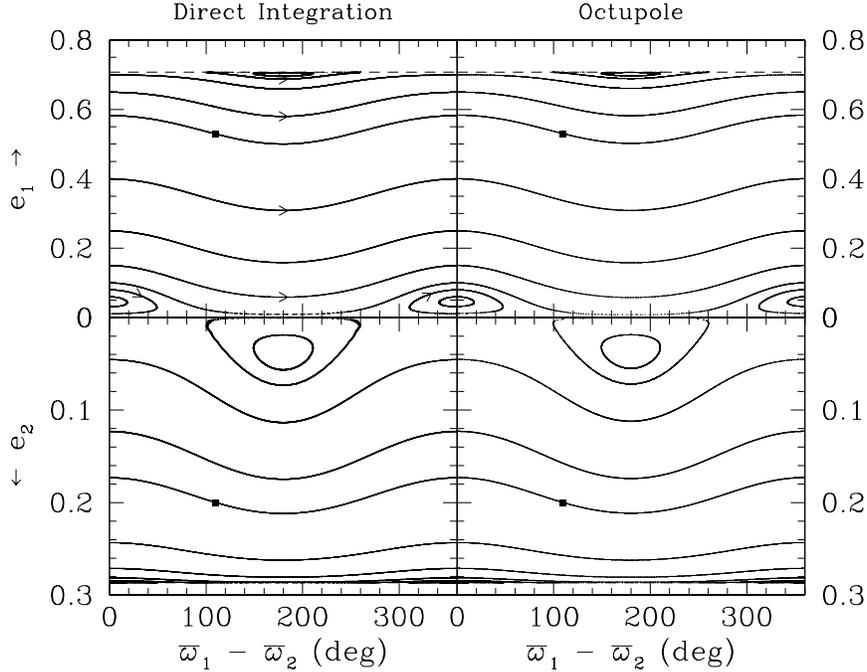}
\caption{
Trajectories in the phase-space diagrams of $e_1$ vs.\ $\dvarpi$ ({\it
upper panels}) and $e_2$ vs.\ $\dvarpi$ ({\it lower panels}) for
two-planet systems with the same masses, initial semimajor axes, and
total angular momentum as the $\sin i = 1$ HD~168443 system.
The left and right panels show the results from direct numerical orbit
integrations and the octupole-level secular perturbation theory,
respectively.
The trajectories through the squares are those of the HD~168443
system, with the squares showing the current parameters of the system.
The initial conditions for the other trajectories are $\dvarpi =
0^\circ$ and $e_1 = 0.06$, $0.08$, $0.10$, $0.15$, $0.25$, $0.40$,
$0.65$, or $0.70$, or $\dvarpi = 180^\circ$ and $e_1 = 0.706$ or
$0.7072$, with $e_2$ being determined from the total angular momentum.
All the direct integrations (except that for the HD~168443 system)
start with the outer planet at apoapse and the inner planet at
opposition.
\label{figure2}}
\end{figure}

To study the effects of $\sin i$, we have performed two direct integrations
of the coplanar HD 168443 system with $\sin i = 0.4$, one starting at
$T_{{\rm peri},1}$ and the other $T_{{\rm peri},2}$.
(Although the two-Kepler fit to the radial velocity observations does not
yield $\sin i_j$, the lack of evidence for stellar wobble in the
{\it Hipparcos} astrometric data limits $\sin i_2 \ga 0.4$ for the outer
planet of HD~168443; Marcy et al. 2001.)
In both cases, the amplitudes of eccentricity oscillations and the
trajectory in $e_1$ (or $e_2$) versus $\dvarpi$ are almost identical to
those shown in Figures 1{\it b} and 2 for $\sin i = 1$, and the factor by
which the eccentricity oscillation period shortens agrees with $\sin i =
0.4$ to better than $\pm 0.001$.
These results are in good agreement with the analytic results derived in
\S~3.3 in the limit $m_1, m_2 \ll m_0$, even though $m_2$ is almost
$44 M_J$ (and $m_2/m_0 \approx 0.042$) when $\sin i = 0.4$.

In addition to the trajectories of the $\sin i = 1$ HD~168443 system,
Figure 2 also shows the trajectories of systems with the same masses,
initial semimajor axes, and total angular momentum as the $\sin i = 1$
HD~168443 system.
(Recall that the octupole results are also valid for other systems with the
same $\beta$, $\lambda$, and $\gamma$.)
There is excellent agreement between the direct-integration and octupole
results in all cases.
The HD~168443 system is not in a secular resonance and its $\dvarpi$
circulates.
Indeed it is far from the relatively small libration islands about the
fixed points at $(\dvarpi, e_1) = (0^\circ, 0.046)$ and
$(180^\circ, 0.702)$ in the phase-space diagram of $e_1$ versus
$\dvarpi$.
The small libration islands and modest eccentricity variations in
Figure 2 are consistent with the fact that these systems have $\lambda
= 0.143$ far from $\lambda_{\rm crit} = 0.836$ for the dimensionless
total angular momentum $\gamma = 0.963$ of these systems.

\begin{figure}[t]
\epsscale{0.7}
\plotone{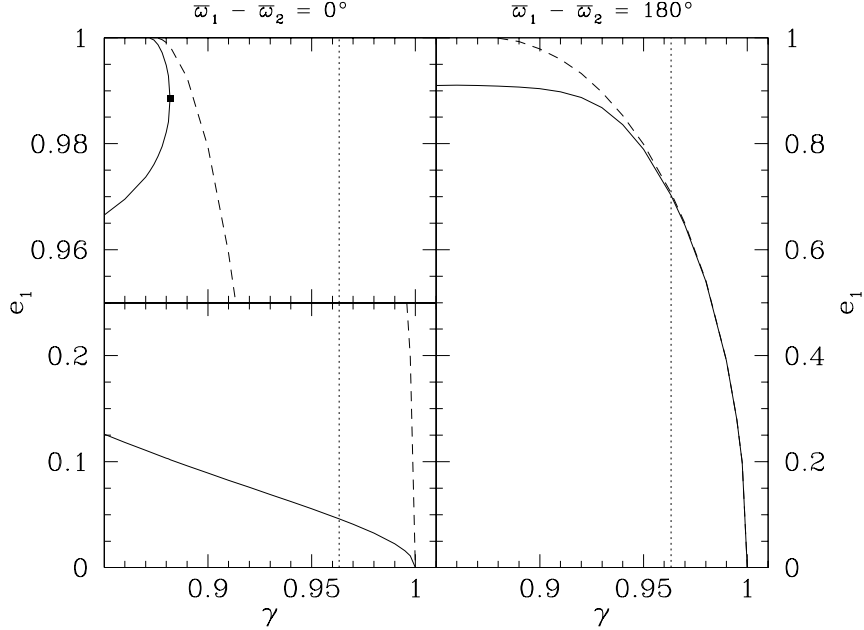}
\caption{
Values of $e_1$ for the fixed points at $\dvarpi = 0^\circ$ ({\it left
panels}) and $180^\circ$ ({\it right panel}) in the phase-space
diagram of $e_1$ vs.\ $\dvarpi$ (like Fig.~2) as a function of the
dimensionless total angular momentum $\gamma = (G_1 + G_2)/(L_1 +
L_2)$, as determined by the octupole theory for systems with the same
$\beta$($=0.126$) and $\lambda$($=0.143$) as the $\sin i = 1$
HD~168443 system.
The dotted lines indicate the dimensionless total angular momentum
$\gamma = 0.963$ of the $\sin i = 1$ HD~168443 system, and the dashed
lines indicate the maximum possible $e_1$ as a function of $\gamma$.
See text for the meaning of the square in the upper left panel.
\label{figure3}}
\end{figure}

The above $e_1$ values for the fixed points were obtained from the
octupole theory as the roots of $d(\dvarpi)/d\tau = 0$ at $\dvarpi =
0^\circ$ and $180^\circ$, with $d(\dvarpi)/d\tau$ from equation
(\ref{dw12dtau}) and $e_2$ from the conservation of total angular
momentum.
As we discussed in \S~3.3, the octupole theory predicts that the fixed
points must be at $\dvarpi = 0^\circ$ or $180^\circ$.
To demonstrate how the octupole theory can be used to explore the
parameter space rapidly, we have used the same root finding procedure
to determine the number and positions of fixed points as a function of
the dimensionless total angular momentum $\gamma = (G_1 + G_2)/(L_1 +
L_2)$ for systems with the same $\beta$ and $\lambda$  as the $\sin i
= 1$ HD~168443 system.
The $e_1$ values for the fixed points at $\dvarpi = 0^\circ$ and
$180^\circ$ as a function of $\gamma$ are shown in the left and right
panels of Figure 3, respectively.
For all values of $\gamma$ shown in Figure 3, there is an elliptic
fixed point at $\dvarpi = 0^\circ$ with relatively small $e_1$ (lower
left panel) and an elliptic fixed point at $\dvarpi = 180^\circ$ with
$e_1$ close to the maximum possible value indicated by the dashed line
(right panel).
These two fixed points are the ones seen above for systems with the
same $\gamma$($= 0.963$; dotted lines in Fig.~3) as the $\sin i = 1$
HD~168443 system.

For $0.872 \la \gamma \la 0.8818$, there are two additional fixed
points at $\dvarpi = 0^\circ$ with $e_1$ very close to $1$ (upper left
panel of Fig.~3).
These additional fixed points emerge at $\gamma \approx 0.8818$ with
the same $e_1$ value (square in the upper left panel of Fig.~3), and
the one with $e_1$ value that increases (decreases) with decreasing
$\gamma$ is an elliptic (hyperbolic) fixed point.
Direct numerical orbit integrations confirm that there are indeed stable
librations about an elliptic fixed point close to the additional one
predicted by the octupole theory.
For example, for $\gamma = 0.88$, the octupole theory
predicts that the additional elliptic fixed point is at $e_1 = 0.9948$ (and
$e_2 = 0.1302$).
A direct integration of a system with these values of $e_1$, $e_2$,
and $\dvarpi$ as
initial conditions (and with the masses and semimajor axes of the
$\sin i = 1$ HD~168443 system and mean anomalies $l_1 = 0^\circ$ and $l_2 =
180^\circ$) shows that this system has $\dvarpi$ librating about $0^\circ$
with an amplitude of about $6^\circ$ and $e_1$ varying between $0.9946$ and
$0.9969$.
It should be noted that the treatment of this system with $e_1 \ge
0.9946$ and $a_1 \approx 0.3\au$ as point masses interacting via
Newtonian gravity is inadequate since the periapse distance of the
inner planet is in fact less than the stellar radius.
However, based on the octupole theory, we expect variants of this system
with much larger $a_1$ (so that the periapse distance of the inner planet
is well outside the stellar radius) but the same $\alpha$ to show similar
libration behaviors.
On the other hand, the inner planet could still come sufficiently close to
the star for general relativistic precession and tidal effects to be
important.
For $\gamma \la 0.872$, the additional elliptic fixed point vanishes,
leaving only the hyperbolic fixed point (upper left panel of Fig.~3).
Both octupole and direct-integration calculations show that systems
with $\gamma \la 0.872$, initial $\dvarpi = 0^\circ$ and initial $e_1$
above the hyperbolic fixed point have $e_1$ increasing to unity in a
finite time.
Again, we expect variants of these systems with larger $a_1$ (so that
the periapse distance of the inner planet is well outside the stellar
radius) but the same $\alpha$ to show similar increase in $e_1$.
However, before $e_1$ reaches unity, either the effects of tides and
general relativistic precession would eventually stop the increase in
$e_1$, or the inner planet would collide with the star.

\subsection{HD~12661}

\begin{figure}[t]
\epsscale{0.5}
\plotone{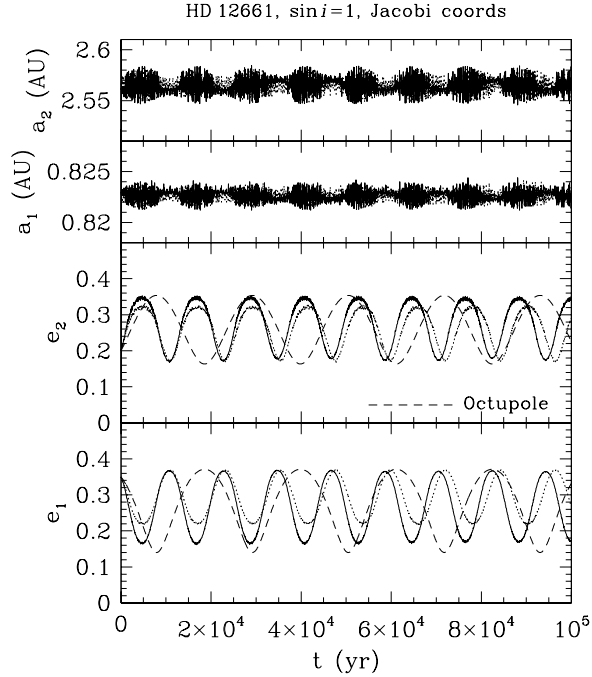}
\caption{
Variations in the Jacobi semimajor axes, $a_1$ and $a_2$,
and eccentricities, $e_1$ and $e_2$, of the $\sin i = 1$ HD~12661
system with the best-fit orbital parameters listed in Table 1.
The solid and dotted lines are from direct integrations starting at
$T_{{\rm peri},1}$ and $T_{{\rm peri},2}$ of Table 1, respectively,
and the dashed lines in the lower two panels are from the octupole
theory.
This system with initial $P_2/P_1 = 0.9975 \times 11/2$ is affected by
the 11:2 mean-motion commensurability, and the direct integration
results are sensitive to the starting epoch.
The direct integration starting at $T_{{\rm peri},1}$ shows irregular
fluctuations in $a_j$ and $e_j$ (note, e.g., the irregular jumps in
the mean values of $a_1$ and $a_2$ at successive minima of $e_2$) and
is most likely chaotic.
The most noticeable effect of the 11:2 commensurability on the direct
integration starting at $T_{{\rm peri},2}$ is the reduction in the
amplitudes of the eccentricity variations.
\label{figure4}}
\end{figure}

The best-fit orbital parameters of the HD~12661 planets from
D. A. Fischer (2002, private communication) and the inferred planetary
masses and semimajor axes for $\sin i = 1$ are listed in
Table \ref{table1}.
Figure 4 shows the variations in the Jacobi $a_j$ and $e_j$, with the
solid and dotted lines from direct numerical orbit integrations starting
at $T_{{\rm peri},1}$ and $T_{{\rm peri},2}$ of Table \ref{table1},
respectively, and the dashed lines in the lower two panels from the
octupole-level secular perturbation theory.
The direct integration results are not consistent with the secular
theory in that the evolution of the orbital elements is sensitive to the
starting epoch.
It turns out that a system with exactly the best-fit orbital parameters
is very close to the 11:2 mean-motion commensurability ($P_2/P_1 =
5.486 = 0.9975 \times 11/2$).
The direct integration starting at $T_{{\rm peri},1}$ (solid lines in
Fig.~4) shows irregular fluctuations in $a_j$ and $e_j$ (note, e.g.,
the irregular jumps in the mean values of $a_1$ and $a_2$ at
successive minima of $e_2$) and is most likely chaotic.
The direct integration starting at $T_{{\rm peri},2}$ (dotted lines in
Fig.~4) does not show any obvious irregular jumps in $a_j$ or $e_j$,
and may be either regular or very weakly chaotic.
But its smaller amplitudes of eccentricity variations are due to the
11:2 commensurability.
The chaos in one (and possibly both) of these calculations is due to
the overlap of the resonances at the 11:2 commensurability (see Holman
\& Murray 1996 and Murray \& Holman 1997 for a similar situation in
the planar elliptic restricted three-body problem).
For both of the direct integrations shown in Figure 4, we have examined
the 10 eccentricity-type mean-motion resonance variables at the 11:2
commensurability and confirmed that some of them alternate between
circulation and libration.
We have extended the direct integrations shown in Figure 4 and found
that both are stable for at least $10^6 \yr$, but we cannot rule out
the possibility that the chaos would lead to instability on longer
timescales.

The orbital period $P_2$ (and the other orbital parameters) of the
outer planet of HD~12661 are not currently known to high precision,
because the time span of the available observations is comparable to
$P_2$.
To study the effects of varying $P_2$, we have performed two sets of
direct integrations with different initial $P_2$, one starting at
$T_{{\rm peri},1}$ and the other at $T_{{\rm peri},2}$.
The initial $P_2$ were chosen such that $(P_2/P_1)/(11/2) = 0.98$,
$0.99$, $1.01$, $1.02$, and $1.03$.
The initial values of the other orbital parameters that can be obtained
from the two-Kepler fit ($P_1$, $K_{1,2}$, $e_{1,2}$, $\omega_{1,2}$,
and $T_{{\rm peri},1,2}$) were fixed at the values listed in
Table \ref{table1}, and the planetary masses and initial semimajor
axes were derived assuming that $\sin i = 1$.
We find that the cases with initial $(P_2/P_1)/(11/2) = 1.01$ are also
affected by the 11:2 mean-motion commensurability while those with
initial $(P_2/P_1)/(11/2) = 0.98$, $0.99$, and $1.02$ show regular
secular evolution.
Although the variations in $a_j$ for the cases with initial $P_2/P_1 =
1.03 \times 11/2 = 0.9997 \times 17/3$ indicate that these cases are
probably affected by the 17:3 commensurability, the variations in
$e_j$ and the trajectories in the diagram of $e_1$ versus $\dvarpi$
are qualitatively indistinguishable from those showing regular secular
evolution (for at least $2 \times 10^5\yr$).

\begin{figure}[t]
\epsscale{0.5}
\plotone{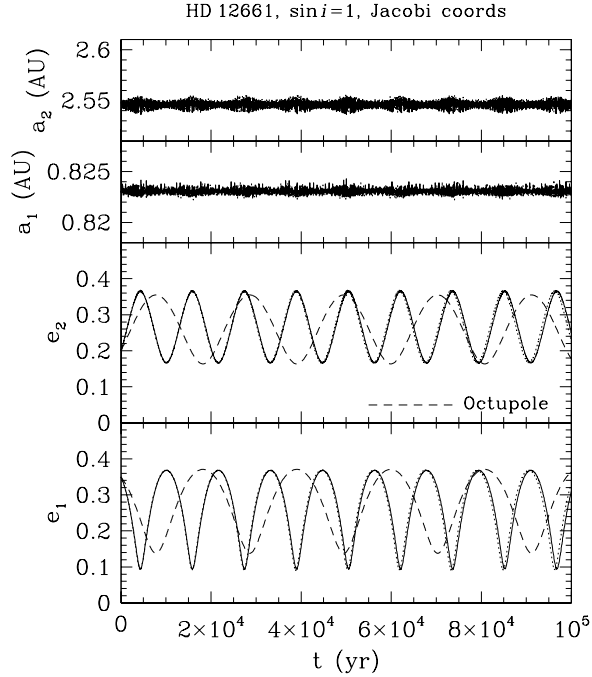}
\caption{
Same as Fig.~4, but for the $\sin i = 1$ HD~12661 system with an
initial $P_2$ such that $P_2/P_1 = 0.99 \times 11/2$.
This system shows regular secular evolution, with the direct
integration results insensitive to the starting epoch.
\label{figure5}}
\end{figure}

\begin{figure}[t]
\epsscale{0.7}
\plotone{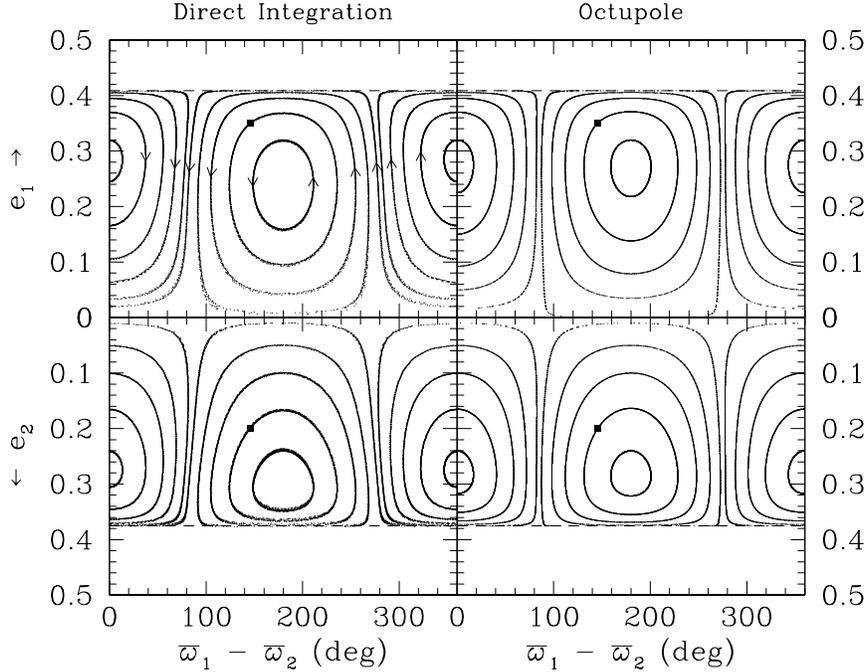}
\caption{
Same as Fig.~2, but for two-planet systems with the same masses,
initial semimajor axes, and total angular momentum as the $\sin i = 1$
HD~12661 system with initial $P_2/P_1 = 0.99 \times 11/2$.
The phase space of these systems with $\lambda$($\approx 0.83$) almost
identical to $\lambda_{\rm crit}$($\approx 0.82$) is dominated by
large libration islands about $\dvarpi = 0^\circ$ and $180^\circ$, and
libration in the secular resonance is almost the only possibility.
The trajectories through the squares are those of the HD~12661 system,
with the squares showing the current parameters of the system.
The initial conditions for the other trajectories are $\dvarpi =
0^\circ$ and $e_2 = 0.01$, $0.05$, $0.10$, $0.165$, or $0.24$, or
$\dvarpi = 180^\circ$ and $e_2 = 0.01$, $0.05$, $0.10$ or $0.24$, with
$e_1$ being determined from the total angular momentum.
All the direct integrations (except that for the HD~12661 system)
start with the outer planet at apoapse and the inner planet at
opposition.
\label{figure6}}
\end{figure}

In Figures 5 and 6 we show examples of the cases with regular secular
evolution.
Figure 5 shows the variations in $a_j$ and $e_j$ for the $\sin i = 1$
HD~12661 system with $P_2/P_1 = 0.99 \times 11/2$.
The solid and dotted lines are from the direct integrations starting at
$T_{{\rm peri},1}$ and $T_{{\rm peri},2}$, respectively, and the dashed
lines are from the octupole-level secular perturbation theory.
The trajectories of this system in the phase-space diagrams of $e_1$
versus $\dvarpi$ and $e_2$ versus $\dvarpi$ are the trajectories marked
by the squares (which denote the current parameters of the HD~12661
system) in Figure 6.
The direct integration results are not sensitive to the starting epoch.
The amplitudes of eccentricity oscillations and the trajectories in the
diagram of $e_1$ (or $e_2$) versus $\dvarpi$ predicted by the octupole
theory are in reasonably good agreement with those from direct
integrations, even though $\alpha$($= 0.323$) is quite large.
The main error of the octupole theory is in the period of eccentricity
oscillations, with the predicted period ($\approx 2.1 \times 10^4\yr$)
about $75\%$ longer than that found in the direct integrations
($\approx 1.2 \times 10^4\yr$).

Figures 5 and 6 show some interesting properties of the HD~12661
system that are shared by all of the cases studied above with
different initial $P_2$, including those affected by the close
proximity to a mean-motion commensurability.
The HD~12661 system is in a secular resonance with $\dvarpi$ librating
about $180^\circ$.
This means that the lines of apsides of the two orbits are on average
anti-aligned.
However, the amplitude of libration of $\dvarpi$ is large, and the
orbital eccentricities of {\it both} planets exhibit large-amplitude
variations.
For the direct integrations shown in Figures 5 and 6, the amplitude of
libration of $\dvarpi$ is $56^\circ$, $e_1$ varies between $0.09$ and
$0.37$, and $e_2$ varies between $0.17$ and $0.37$.
HD~12661 is the first extrasolar planetary system found to have
$\dvarpi$ librating about $180^\circ$.
The outer two planets of the $\upsilon$~And system (Rivera \&
Lissauer 2000; Lissauer \& Rivera 2001; Chiang et al. 2001) and the
two planets about GJ~876 (Laughlin \& Chambers 2001; Lee \& Peale
2002) are also likely in the secular resonance involving $\dvarpi$,
but in both of these cases $\dvarpi$ librates about $0^\circ$.
(In the case of GJ~876, the secular resonance is associated with the
simultaneous librations of both lowest order mean-motion resonance
variables at the 2:1 commensurability.)

The libration of $\dvarpi$ and the large-amplitude variations of both
$e_j$ in the HD~12661 system were anticipated by the analytic results
derived in \S~3.3.
As we discussed in \S~3.3, the octupole theory predicts that large
libration islands and large-amplitude variations of both $e_j$ are
likely if $\lambda = L_1/L_2 \approx \lambda_{\rm crit} = 2
\gamma^2/(5 - 3\gamma^2)$.
The HD~12661 system has $\lambda \approx 0.83$, which is almost
identical to $\lambda_{\rm crit} \approx 0.82$ for the dimensionless
total angular momentum $\gamma \approx 0.96$ of this system.
In addition to the trajectories of the $\sin i = 1$ HD~12661 system
with $P_2/P_1 = 0.99 \times 11/2$, Figure 6 also shows the
trajectories of systems with the same masses, initial semimajor axes,
and total angular momentum.
The direct-integration results (left panels) are in reasonably good
agreement with the octupole results (right panels) and confirm that
the phase space is dominated by large libration islands about $\dvarpi
= 0^\circ$ and $180^\circ$.
The only discrepancy is a narrow region of circulation represented by
a single trajectory in the direct-integration results.
Thus libration of $\dvarpi$ is almost the only possibility.
We can see in Figure 6 that $e_1$ decreases from near its maximum
possible value to near zero and $e_2$ increases from near zero to near
its maximum possible value at $\dvarpi \approx 90^\circ$, and vice
versa at $\dvarpi \approx 270^\circ$, again in agreement with the
analysis in \S~3.3.

Kiseleva-Eggleton et al. (2002) have recently reported that the $\sin
i = 1$ HD~12661 system with the best-fit orbital parameters in
Table \ref{table1} is regular based on the MEGNO (Mean Exponential
Growth of Nearby Orbits) technique.
However, they assumed that the best-fit orbital parameters are in
astrocentric coordinates and showed results from only one
(unspecified) starting epoch.
We have repeated the direct integrations shown in Figure 4, but with
the best-fit orbital parameters assumed to be in astrocentric
coordinates. 
Because of the larger fluctuations in the astrocentric $a_2$, the
average ratio of the orbital periods is more sensitive to the starting
epoch.
The integration starting at $T_{{\rm peri},1}$ turns out to be
sufficiently far from the 11:2 commensurability that it shows regular
secular evolution similar to those in Figure 5.
The integration starting at $T_{{\rm peri},2}$ is qualitatively
similar to the integration in Jacobi coordinates starting at the same
epoch (dotted lines in Figure 4) and is thus either regular or very
weakly chaotic.
These results again demonstrate the importance of
interpreting the orbital parameters from multiple Kepler fits as
orbital parameters in Jacobi coordinates.

The best-fit orbital parameters of the HD~12661 planets in
Table \ref{table1} were provided by D.~A.~Fischer (2002, private
communication), and some of them differ slightly from those published
later in Fischer et al. (2003).
Adopting the parameters in Fischer et al. (2003) does not
qualitatively change the results presented here, although
the smaller $\dvarpi$ for the parameters in Fischer et al. (2003) does
lead to slightly larger amplitude of libration of $\dvarpi$ and
slightly larger variations in $e_j$.

After this paper was completed, we learned of contemporaneous work on
the HD~12661 system by Go\'zdziewski (2003) and Go\'zdziewski \&
Maciejewski (2003).
Go\'zdziewski (2003) has examined a wide region of the parameter
space near the HD~12661 best-fit solution in Table \ref{table1}, using
astrocentric coordinates and a combination of the MEGNO technique and
direct integrations.
His numerical results also show the proximity of the best-fit solution
to the 11:2 commensurability and the robustness of the libration of
$\dvarpi$.
Go\'zdziewski \& Maciejewski (2003) have subsequently reported a
best-fit solution to the data set published by Fischer et al. (2003)
with $P_2 \approx 1660\, $days or $P_2/P_1 \approx 6.3$.
This illustrates the uncertainties mentioned above in the orbital
parameters of the outer planet due to the time span of the available
observations being comparable to $P_2$.
Their best-fit solution also shows libration of $\dvarpi$ about
$180^\circ$ and large-amplitude variations of both $e_j$.

Mayor et al. (2003) have reported the detection of two planets around the
star HD~83443.
There is considerable doubt about the existence of the outer planet in
this system (Butler et al. 2002), but it is interesting to note that
the system as reported by Mayor et al. shares some of the properties of the
HD~12661 system.
The best-fit orbital periods of the HD~83443 planets are nearly in the
ratio 10:1, and direct integrations of this system (neglecting the
effects of tides and general relativistic precession) show that its
evolution is also affected by the close proximity to the mean-motion
commensurability.
The HD~83443 system also shows large-amplitude variations of both
eccentricities, although its $\dvarpi$ is circulating.
The value of $\lambda$($\approx 0.96$) of the HD~83443 system is not
as close to its value of $\lambda_{\rm crit}$($\approx 0.79$) as in
the HD~12661 system, and there is a larger region of circulation in
the phase space of systems with the same masses, initial semimajor
axes, and total angular momentum as the HD~83443 system.
It should be noted that the inner planet of the HD~83443 system is
sufficiently close to the star ($P \approx 3\,$days) that tides and
general relativistic precession can significantly affect the dynamics
of the system (Wu \& Goldreich 2002).

\subsection{Small Mutual Inclination}

To study the effects of a small mutual inclination angle $i_{\rm mu}$
between the orbits, we have repeated the direct integrations of the
HD~168443 and HD~12661 systems shown in Figures 1{\it b} and 5 with
initial $i_{\rm mu} = 1^\circ$ or $2^\circ$.
We assumed that the intersection of the orbital planes is initially
along the line of sight, so that, initially, $\sin i_1 = \sin i_2 = 1$
and the mutual inclination is the difference in the longitudes of
ascending node referenced to the plane of the sky:
$i_{\rm mu} = \Omega_2 - \Omega_1$.
The results are nearly identical to those with coplanar orbits,
confirming the analysis in \S~3.1 based on an expansion of the more
general form of the octupole theory in powers of $i_{\rm mu}$.

\subsection{Stability}

The applicability of the octupole theory is limited by its use of an
expansion in $r_1/r_2$ and orbit averaging.
In particular, the orbit averaging eliminates the effects of mean-motion
commensurabilities and the possible development of instabilities.
The direct integrations reported in \S~4.1 show that the coplanar HD~168443
system is stable for $\sin i \ga 0.4$, and Marcy et al. (2001) have
found that this system is stable for even smaller $\sin i$.
The stability of the HD~168443 system is consistent with the empirical
stability criteria for hierarchical triple systems derived most recently by
Eggleton \& Kiseleva (1995) and Mardling \& Aarseth (2001).
For example, the $\sin i = 1$ HD~168443 system has $\alpha = a_1/a_2 =
0.102$, while the stability criterion of Eggleton \& Kiseleva requires
initial $a_2 (1 - e_2)/[a_1 (1 + e_1)] \ga 1.65$ or $\alpha \la 0.316$
and that of Mardling \& Aarseth requires initial $a_2 (1 - e_2)/a_1
\ga 3.17$ or $\alpha \la 0.252$.
The criterion of Eggleton \& Kiseleva was tested over a wide range of mass
ratios (including ones similar to those of the HD~168443 system) and is
expected to be reliable to about $20\%$, while the criterion of Mardling \&
Aarseth does not have explicit dependence on $m_1/m_0$ and may not be very
accurate for $m_1/m_0$ significantly different from unity.
For a more detailed comparison with the above stability criteria, we have
performed a set of direct integrations with different initial $a_2$:
$a_2 = 1.299$--$0.738 \au$ or $\alpha = 0.227$--$0.400$.
The other initial conditions are identical to those of the $\sin i = 1$
calculation shown as the solid lines in Figure 1{\it b}.
The evolutions of $\alpha$ are shown in Figure 7.
Except for the case with initial $\alpha = 0.333$, which is apparently
stabilized by the 5:1 mean-motion commensurability, there is a clear
boundary between stable and unstable configurations at initial $\alpha
\approx 0.30$.
We do not expect this stability boundary to move down significantly if we
increase the integration length, since $10^5 \yr$ span many eccentricity
oscillation cycles (if the orbits are stable) and, as we can see in Figure 7,
instabilities usually manifest themselves on much shorter timescales.
The stability criterion of Eggleton \& Kiseleva is a few percent
larger than the stability boundary determined here, consistent with
the $\approx 20\%$ accuracy of the former.
Even the stability criterion of Mardling \& Aarseth is only $16\%$ smaller
than the stability boundary determined here (see, however, the next
paragraph).
Finally, we note that some of the configurations in Figure 7 that are
stable for at least $10^5\yr$ appear to be chaotic (e.g., the case
with initial $\alpha = 1/4$).
As in the case of the HD~12661 system, the chaos is due to
the close proximity to a high-order mean-motion commensurability (8:1
for the case with initial $\alpha = 1/4$).

\begin{figure}[t]
\epsscale{0.8}
\plotone{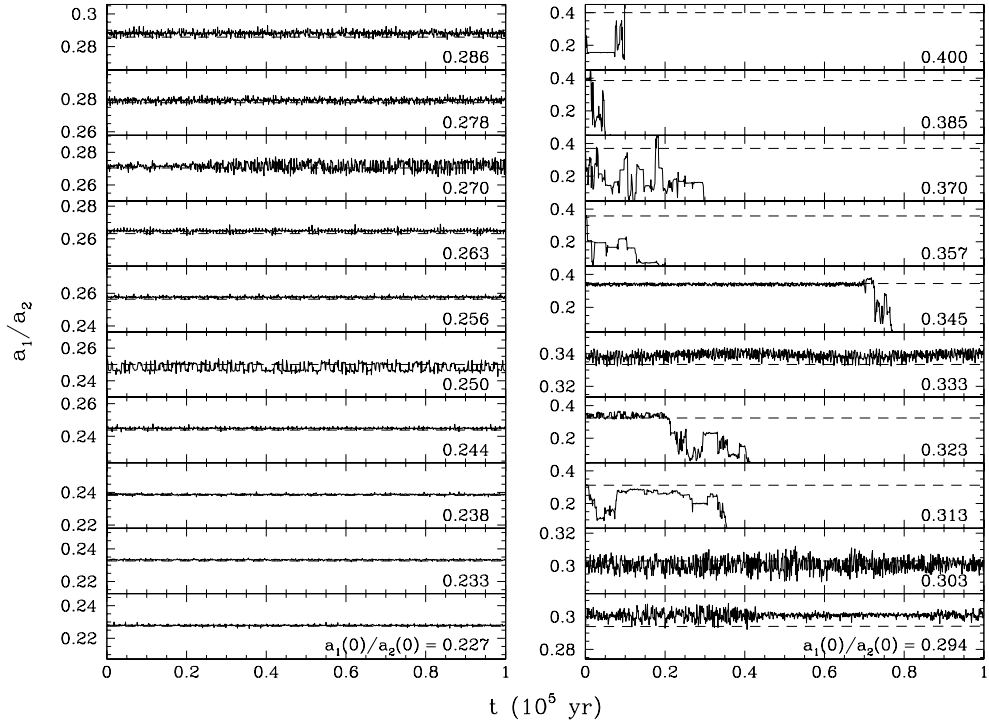}
\caption{
Evolutions of $a_1/a_2 = \alpha$ for a set of direct integrations
with different initial $a_2$: $a_2 = 1.299$--$0.738 \au$ or $\alpha =
0.227$--$0.400$.
The other initial conditions are identical to those of the $\sin i =
1$ HD~168443 calculation shown as the solid lines in Fig.~1{\it b}.
The dashed line in each panel indicates the initial $\alpha$.
Some of the cases that are stable for at least $10^5 \yr$ are affected
by the close proximity to a high-order mean-motion commensurability
(e.g., 8:1 for the case with initial $\alpha = 1/4$) and appear to be
chaotic.
The case with initial $\alpha = 0.333$ is apparently stabilized by the
5:1 commensurability.
\label{figure7}}
\end{figure}

\begin{figure}[t]
\epsscale{0.8}
\plotone{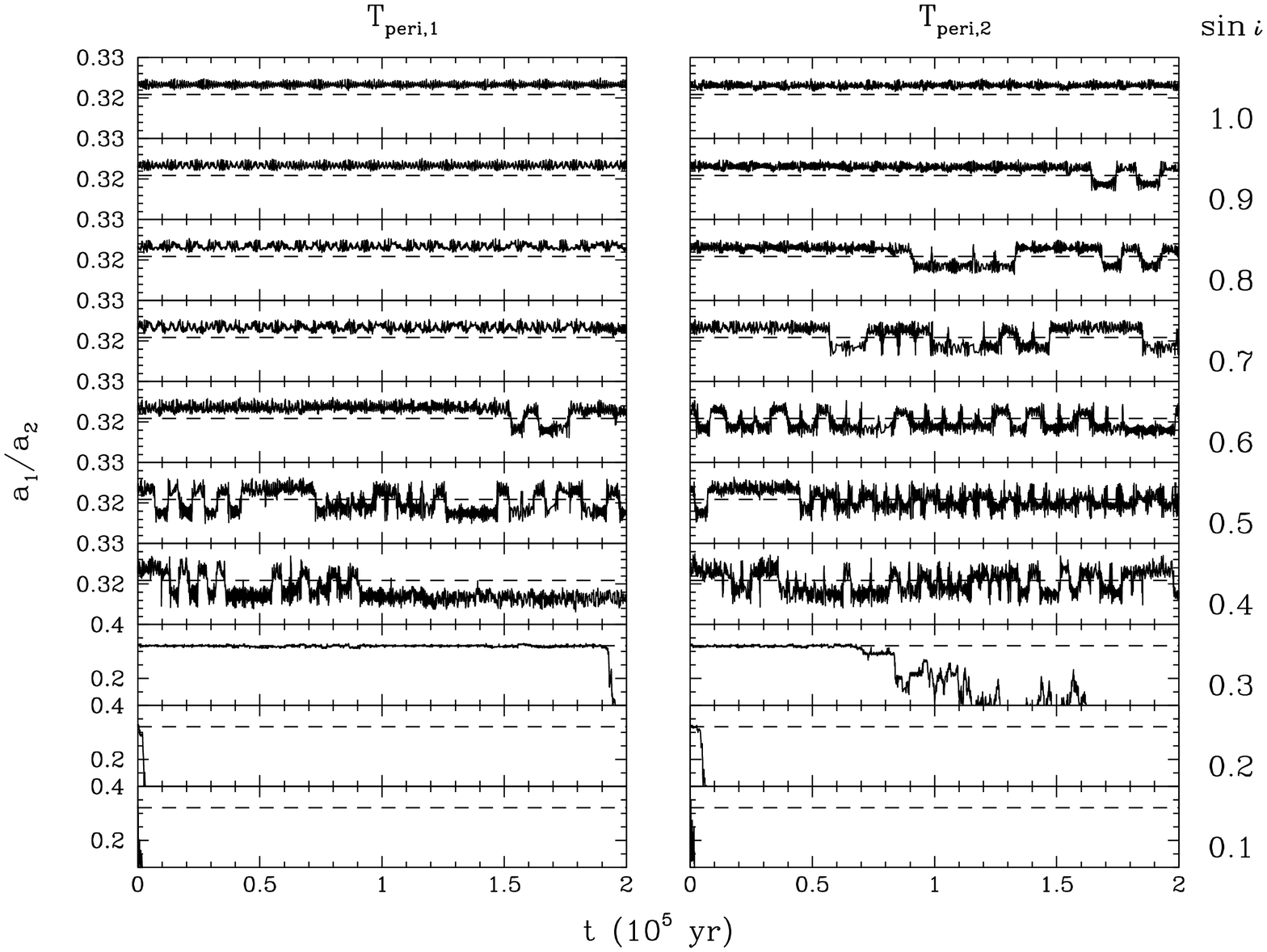}
\caption{
Evolutions of $a_1/a_2 = \alpha$ for the HD~12661 system with
$P_2/P_1 = 0.99 \times 11/2$ and $\sin i = 0.1$, $0.2, \ldots, 1$.
The initial values of $P_j$, $K_j$, $e_j$, $\omega_j$, and
$T_{{\rm peri},j}$ are the same as those of the system shown in
Fig.~5, and the planetary masses and initial semimajor axes were
derived for the assumed $\sin i$.
The left and right panels are from direct integrations starting at
$T_{{\rm peri},1}$ and $T_{{\rm peri},2}$, respectively.
The dashed line in each panel indicates the location of the 11:2
mean-motion commensurability.
\label{figure8}}
\end{figure}

In \S~4.2 we have considered the coplanar HD~12661 system with $\sin i
= 1$ (i.e., minimum planetary masses) only.
Unlike the HD~168443 system, which has $\alpha = a_1/a_2 \sim 0.1$ and is
stable for any reasonable value of $\sin i$, the HD~12661 system has
$\alpha \sim 0.32$ (which is close to the stability boundary for the
systems shown in Fig.~7 with the planetary masses of the $\sin i = 1$
HD~168443 system) and could be unstable for moderately small $\sin i$.
To study the effects of $\sin i$, we have performed two sets of direct
integrations with different $\sin i$.
The initial values of $P_j$, $K_j$, $e_j$, $\omega_j$, and
$T_{{\rm peri},j}$ are the same as those of the HD~12661 system with
$P_2/P_1 = 0.99 \times 11/2$ shown in Figure 5, and the planetary
masses and initial semimajor axes were derived for the assumed $\sin i$.
The evolutions of $\alpha$ are shown in Figure 8 for $\sin i = 0.1$,
$0.2, \ldots, 1$.
The left and right panels are from the direct integrations starting at
$T_{{\rm peri},1}$ and $T_{{\rm peri},2}$, respectively.
The dashed line in each panel indicates the location of the 11:2
mean-motion commensurability.
The influence of the 11:2 commensurability increases with decreasing
$\sin i$ (which is expected since the width of a resonance increases
with increasing planetary masses), but the cases with $\sin i > 0.3$
are stable (for at least $2 \times 10^5\yr$).
All of the cases with $\sin i \la 0.3$ are unstable.
Are the stability criteria of Eggleton \& Kiseleva (1995) and Mardling
\& Aarseth (2001) consistent with this transition from stable to
unstable configurations at $\sin i \approx 0.3$?
For the configurations shown in Figure 8, which have the same initial
$P_1$ and $P_2$, the initial $\alpha$ decreases slightly from $0.323$
to $0.322$ as $\sin i$ decreases from $1$ to $0.1$ (and $m_1$ and
$m_2$ increase; see eqs. [\ref{period1}] and [\ref{period2}]).
For comparison, the stability boundary predicted by the criterion of
Eggleton \& Kiseleva decreases from initial $\alpha = 0.444$ to
$0.348$ as $\sin i$ decreases from $1$ to $0.1$, with initial $\alpha
= 0.397$ at $\sin i = 0.3$.
This stability boundary is consistent with the transition at $\sin i
\approx 0.3$ if we simply move it down by $23\%$, which is compatible
with the $\approx 20\%$ accuracy of the stability criterion of
Eggleton \& Kiseleva.
On the other hand, the stability boundary predicted by the criterion
of Mardling \& Aarseth changes only slightly from initial $\alpha =
0.254$ to $0.253$ as $\sin i$ decreases from $1$ to $0.1$.
Thus, this stability boundary is not consistent with the transition at
$\sin i \approx 0.3$, unless we allow adjustments that change with
$\sin i$.
This is not surprising since the stability criterion of Mardling \&
Aarseth does not have explicit dependence on $m_1/m_0$ and is not
expected to be very accurate for $m_1/m_0$ significantly different
from unity.

\section{CONCLUSIONS}

We have investigated the dynamical evolution of coplanar hierarchical
two-planet systems where the ratio of the orbital semimajor axes
$\alpha = a_1/a_2$ is small.
Hierarchical two-planet systems are common among the known multiple
planet systems and are likely to be common in the overall population
of extrasolar planetary systems.
We began by showing that the orbital parameters obtained from a two
(or more generally multiple) Kepler fit to the radial velocity
variations of a host star are best interpreted as Jacobi coordinates
and that these coordinates should be used in any analyses of
hierarchical (and possibly other types of) planetary systems.
We used the HD~168443 system as an example to show that the use of
astrocentric coordinates can introduce significant high-frequency
variations in orbital elements that should be nearly constant on
orbital timescales, leading to erroneous sensitivity of the evolution
of the orbital elements on the starting epoch.

An approximate theory that can be applied to hierarchical two-planet
systems with a wide range of masses, orbital eccentricities, and
inclinations is the octupole-level secular perturbation theory, which
was developed for general hierarchical triple systems by Marchal
(1990), Krymolowski \& Mazeh (1999), and Ford et al. (2000).
The octupole theory is based on an expansion to order $\alpha^3$ and
orbit-averaging.
We showed that the octupole approximation reduces the coplanar problem
to one degree of freedom, with $e_1$ (or $e_2$) and $\dvarpi$ as the
relevant phase-space variables.
An analysis of the octupole equations yielded the following results:
(1) the scaling properties of the equations imply that the amplitudes
of eccentricity oscillations and the trajectory in the phase-space
diagram of $e_1$ (or $e_2$) versus $\dvarpi$ should be independent of
the inclination $i$ of both orbits from the plane of the sky (as long
as $m_1, m_2 \ll m_0$) and that the period of eccentricity
oscillations should be proportional to $\sin i$;
(2) the fixed points in the phase-space diagram must be at $\dvarpi =
0^\circ$ or $180^\circ$; and
(3) if the ratio of the maximum orbital angular momenta, $\lambda =
L_1/L_2 \approx (m_1/m_2) \alpha^{1/2}$, for given semimajor axes is
approximately equal to $\lambda_{\rm crit} = 2 \gamma^2/(5 -
3\gamma^2)$, where $\gamma = (G_1 + G_2)/(L_1 + L_2)$ is the
dimensionless total angular momentum, then libration of $\dvarpi$ is
almost certain, with possibly large amplitude variations of both
eccentricities.

We used both the octupole-level secular perturbation theory and direct
numerical orbit integrations to study the dynamical evolution of the
HD~168443 and HD~12661 systems and their variants.
The HD~168443 system is not in a secular resonance and its $\dvarpi$
circulates.
For the family of systems with the same masses, initial semimajor
axes, and total angular momentum as the $\sin i = 1$ HD~168443 system,
there are two relatively small libration islands in the phase-space
diagram of $e_1$ (or $e_2$) versus $\dvarpi$, one about a fixed point
at $\dvarpi = 0^\circ$ and the other about a fixed point at $\dvarpi =
180^\circ$, and the trajectory of the HD~168443 system is far from
both.
In all cases, the octupole results are in excellent agreement with the
direct-integration results.
Direct integrations of the HD~168443 system with $\sin i = 0.4$ also
confirmed the analytic octupole results on the effects of $\sin i$.
To demonstrate how the octupole theory can be used to explore the
parameter space rapidly, we used the octupole theory to determine the
number and positions of fixed points as a function of the total
angular momentum for systems with the same masses and semimajor axes
as the $\sin i = 1$ HD~168443 system.
For sufficiently low total angular momentum, the existence of fixed
point(s) in addition to those appropriate for the total angular
momentum of the HD~168443 system leads to systems with unusual
behaviors such as $e_1$ increasing to unity in a finite time or
libration of $\dvarpi$ about $0^\circ$ with $e_1$ very close to 1.
These results were again confirmed by direct integrations.
Direct integrations of systems similar to the $\sin i = 1$ HD~168443
system, but with different initial $a_2$, showed that these systems
with relatively massive planets [$(m_1 + m_2)/m_0 \approx 0.024$] are
unstable if the initial $\alpha \ga 0.30$.

The HD~12661 system is the first extrasolar planetary system found to
have $\dvarpi$ librating about $180^\circ$.
The secular resonance means that the lines of apsides of the two
orbits are on average anti-aligned, although the amplitude of
libration of $\dvarpi$ is large.
The libration of $\dvarpi$ and the large-amplitude variations of both
orbital eccentricities in the HD~12661 system, which has $\lambda
\approx 0.83$ and $\lambda_{\rm crit} \approx 0.82$, are consistent
with the analytic results on systems with $\lambda \approx
\lambda_{\rm crit}$.
Direct integrations showed that the evolution of the $\sin i = 1$
HD~12661 system with the best-fit orbital parameters ($P_2/P_1
= 0.9975 \times 11/2$) is affected by the close proximity to the 11:2
mean-motion commensurability, but that small changes in the orbital
period of the outer planet within the uncertainty can result in
configurations that are not affected by mean-motion
commensurabilities.
For the family of systems with the same masses, initial semimajor
axes, and total angular momentum as the $\sin i = 1$ HD~12661 system
with $P_2/P_1 = 0.99 \times 11/2$, which are not affected by
mean-motion commensurabilities, the octupole results are in reasonably
good agreement with the direct-integration results, even though
$\alpha$($\approx 0.32$) is quite large.
The phase space of this family of systems with $\lambda \approx
\lambda_{\rm crit}$ is dominated by large libration islands about
$\dvarpi = 0^\circ$ and $180^\circ$, which again confirmed the
analytic results on systems with $\lambda \approx \lambda_{\rm crit}$,
and libration in the secular resonance is almost the only possibility.
Finally, we showed that the HD~12661 system is unstable if $\sin i \la
0.3$ [or $(m_1 + m_2)/m_0 \ga 0.012$].

The comparisons with direct numerical orbit integrations established
that the octupole-level secular perturbation theory is highly accurate
for coplanar hierarchical two-planet systems with $\alpha \la 0.1$ and
reasonably accurate even for systems with $\alpha$ as large as $1/3$,
provided that $\alpha$ is not too close to a significant mean-motion
commensurability or above the stability boundary.
We have focused our study in this paper on coplanar systems in which
the star and the planets can be treated as point masses interacting
via Newtonian gravity, where a small mutual inclination between the
orbits is shown to have little effect on the secular evolution.
The more general form of the octupole theory is applicable to systems
with arbitrary mutual inclinations.
If the inner or both planets are very close to the star, tidal
interactions and general relativistic precession can be important
(see, e.g., Wu \& Goldreich 2002).
Nagasawa et al. (2003) have also discussed the effects of additional
orbital precessions due to interactions with the protoplanetary disk
during the epoch of disk depletion.
It is straightforward to modify the octupole theory to include any of
these effects (see, e.g., Blaes et al. 2002 for relativistic
precession).

\acknowledgments
We thank Debra Fischer for furnishing the orbital parameters of the
HD~12661 system before publication.
We also thank O. Blaes, G. W. Marcy, N. Murray, E. J. Rivera, and Y. Wu for
informative discussions and an anonymous referee for comments that
greatly improved the presentation.
This research was supported in part by NASA grants NAG5~11666 and
NAG5~12087.

\end{document}